\documentclass{ieeeaccess}

\usepackage{amsmath,amssymb,amsfonts}
\usepackage[cmintegrals]{newtxmath}
\usepackage{algorithmic}
\usepackage{graphicx}
\usepackage{textcomp}

\usepackage{microtype}

\usepackage[utf8]{inputenc}
\usepackage{url}
\usepackage[sort]{cite}
\DeclareGraphicsExtensions{.pdf,.jpeg,.png}
\usepackage{array,multirow}
\newcolumntype{P}[1]{>{\centering\arraybackslash}p{#1}}
\usepackage{caption}
\usepackage{subcaption}
\usepackage[export]{adjustbox}
\usepackage{cleveref}
\crefformat{section}{§#2#1#3}
\usepackage{tabularx,booktabs}
\usepackage{array, makecell}
\setcellgapes{3pt}
\newcommand{\PreserveBackslash}[1]{\let\temp=\\#1\let\\=\temp}
\newcolumntype{Y}{>{\centering\arraybackslash}X}
\newcolumntype{C}[1]{>{\PreserveBackslash\centering}p{#1}}

\def\BibTeX{{\rm B\kern-.05em{\sc i\kern-.025em b}\kern-.08em
    T\kern-.1667em\lower.7ex\hbox{E}\kern-.125emX}}
    
\begin{document}

\history{Received May 2, 2019, accepted May 28, 2019, date of publication xxxx 00, 0000, date of current version May 29, 2019.}
\doi{10.1109/ACCESS.2019.2920100}

\title{Atomic-SDN: Is Synchronous Flooding the Solution to Software-Defined Networking in IoT?}

\author{
\uppercase{Michael Baddeley\authorrefmark{1},
Usman Raza\authorrefmark{2}, 
Aleksandar Stanoev\authorrefmark{2}, 
George Oikonomou\authorrefmark{1}, 
Reza Nejabati\authorrefmark{1}, 
Mahesh Sooriyabandara\authorrefmark{2}, 
and Dimitra Simeonidou}.\authorrefmark{1}}

\address[1]{Department of Electronic and Electrical Engineering, University of Bristol, Bristol, BS8 1TH, UK (e-mail: m.baddeley; reza.nejabati; g.oikonomou; dimitra.simeonidou@bristol.ac.uk)}

\address[2]{Toshiba Research Europe Ltd., Bristol, BS1 4ND, UK (email: usman.raza; aleksandar.stanoev, mahesh@toshiba-trel.com)}

\tfootnote{The authors wish to acknowledge the financial support of the Engineering and Physical Sciences Research Council (EPSRC) Centre for Doctoral Training (CDT) in Communications (EP/I028153/1), as well as Toshiba Research Europe Ltd.}

\markboth
{M. Baddeley \headeretal: Atomic-SDN: Is Synchronous Flooding the Solution to SDN in IoT?}
{M. Baddeley \headeretal: Atomic-SDN: Is Synchronous Flooding the Solution to SDN in IoT?}

\corresp{Corresponding author: Michael Baddeley (e-mail: m.baddeley@bristol.ac.uk).}

\begin{abstract}
The adoption of Software Defined Networking (SDN) within traditional networks has provided operators the ability to manage diverse resources and easily reconfigure networks as requirements change. Recent research has extended this concept to IEEE 802.15.4 low-power wireless networks, which form a key component of the Internet of Things (IoT). However, the multiple traffic patterns necessary for SDN control makes it difficult to apply this approach to these highly challenging environments. This paper presents Atomic-SDN, a highly reliable and low-latency solution for SDN in low-power wireless. Atomic-SDN introduces a novel Synchronous Flooding (SF) architecture capable of dynamically configuring SF protocols to satisfy complex SDN control requirements, and draws from the authors' previous experiences in the IEEE EWSN Dependability Competition: where SF solutions have consistently outperformed other entries. Using this approach, Atomic-SDN presents considerable performance gains over other SDN implementations for low-power IoT networks. We evaluate Atomic-SDN through simulation and experimentation, and show how utilizing SF techniques provides latency and reliability guarantees to SDN control operations as the local mesh scales. We compare Atomic-SDN against other SDN implementations based on the IEEE 802.15.4 network stack, and establish that Atomic-SDN improves SDN control by orders-of-magnitude across latency, reliability, and energy-efficiency metrics.
\end{abstract}


\titlepgskip=-15pt

\maketitle
\pubid{\begin{minipage}[t]{\textwidth}\ \\[4pt]
        \centering\tiny{\copyright 2019 IEEE.  Personal use of this material is permitted.  Permission from IEEE must be obtained for all other uses, in any current or future media, including reprinting/republishing this material for advertising or promotional purposes, creating new collective works, for resale or redistribution to servers or lists, or reuse of any copyrighted component of this work in other works.}
\end{minipage}} 

\section{Introduction}
\label{sec_intro}
\noindent\textbf{Context and Challenge:} 
By providing an abstracted network view and virtualized network functions, Software Defined Networking (SDN) allows services to be centrally programmed onto functionally agnostic hardware. The ability to reconfigure the network as needed, quickly install new protocols, or slice network resources across applications and tenants, allows networks to adapt to changing requirements or shifts in business needs. 

Implementation of this concept has been immensely successful in data-centre and radio-access networks \cite{sdn_comprehensive_survey}, where reliable and low-latency links support communication between a logically centralized controller (or a number of controllers) and the rest of the network. This ensures that control tasks, such as installing flowtable rules, cause minimum disruption to the data forwarding operations.

Extending the benefits of SDN to the wireless domain and to low-power sensor networks has gained popularity in recent years \cite{sdwn_opportunities_challenges}. Consequently, a number of SDN architectures have been implemented for IEEE 802.15.4 low-power wireless networks. This has provided a model for SDN control and configuration across a low-power multi-hop mesh. However the shared nature of the underlying wireless medium, a need to transmit data over multi-hops, and stringent resource constraints pose significant challenges not present in the conventional wired SDN networks.

\begin{figure*}[ht]
  \centering
  \includegraphics[width=1.0\textwidth]{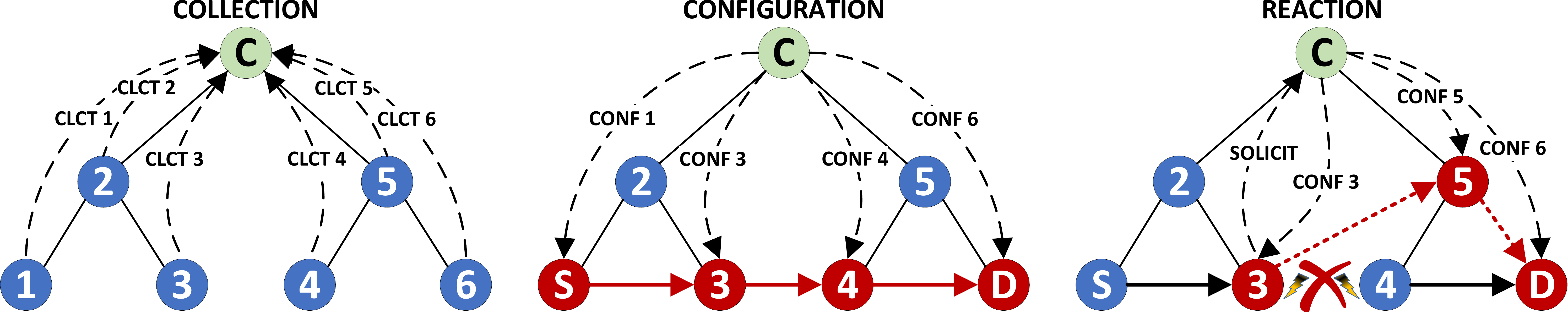}
  \caption{Core SDN services within a low-power mesh network: \textit{Collection} (CLCT), \textit{Configuration} (CONF), and \textit{Reaction} (SOLICIT + CONF). Nodes that need to receive instruction from the SDN controller are marked in red. Nodes \textit{S} and \textit{D} mark the source and destination nodes for a point to point link across the mesh.} 
  \label{fig:sdn_all_opprtunities}
\end{figure*}

Firstly, additional SDN overhead increases contention over the shared wireless medium as well as competition with existing network protocols. Recent efforts have attempted to address this issue through optimization of the SDN protocols, reduction in message frequency, and prioritization or dedication of network resources \cite{sdwn,sdn-wise,coral-sdn,baddeley_isolating_sdn_control_2017}. Secondly, the multi-hop mesh topology prevalent in low-power wireless networks introduces delay and unreliability at each hop. This motivates the need for an SDN protocol that supports an ultra-fast hop-by-hop forwarding scheme and diversity techniques to achieve a very high reliability. Lastly, SDN requires frequent back-and-forth communication between the controller(s) and network nodes. The flow of this traffic follows a variety of different patterns including \textit{many-to-one}, \textit{one-to-many} and \textit{one-to-one} communication. Unfortunately, the standard protocols such as the IPv6 Routing Protocol for Low Power and Lossy Networks (RPL), employed by several wireless SDN architectures, provide less-than-optimal performance for the plurality of these traffic patterns. 

A radically different design is required, supporting all the traffic patterns to provide the following three essential core SDN services, which are presented in Figure \ref{fig:sdn_all_opprtunities}. 

\begin{itemize}
\item \textit{Collection (many-to-one)}: allow the controller to gather node information and infer the current network state. 
\item \textit{Configuration (one-to-many}): allow the controller to provide instruction to all nodes within the mesh, or a subset thereof.
\item \textit{Reaction (one-to-one + one-to-many)}: allow nodes to query or alert the controller for instructions on how to react to new input or events.
\end{itemize}

\noindent This new approach ultimately needs to remove the complexities of mapping traditional SDN architecture to the currently available protocols in low-power wireless; where fundamental challenges arise from the controller not only having to communicate reliably with all nodes, but that each individual operation (for example, to set a path between two nodes) can mean the replication of control messages across multiple nodes in order to correctly configure the network. 

This paper therefore proposes utilizing Synchronous Flooding (SF) as a basis for SDN control in IEEE 802.15.4 low-power wireless networks, and draws on the authors' extensive experience in implementing SF solutions for the International Conference on Embedded Wireless Systems and Networks (EWSN) Dependability Competition \cite{raza_competition_2018, raza_competition_2017}, where a version of Atomic-SDN placed 2$^{nd}$ for both \textit{collection} and \textit{dissemination} categories in 2019 \cite{baddeley_competition_2019}.

\begin{figure*}[ht]
  \centering
  \includegraphics[width=1\textwidth]{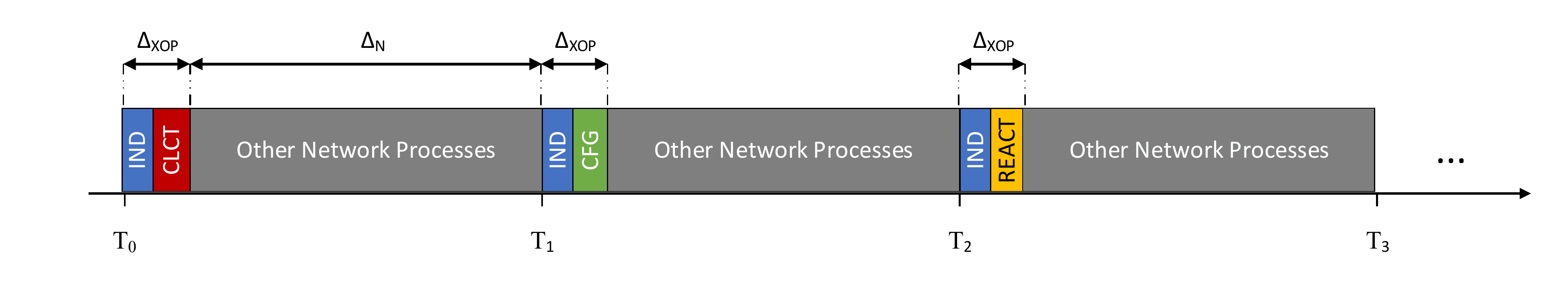}
  \caption{Atomic-SDN uses time-sliced SF control to maximize network resource utilization during control periods. SDN \textit{collection} (CLT), \textit{configuration} (CFG), and \textit{reaction} (REACT) opportunities are preceded by a \textit{indication} (IND) flood.} 
  \label{fig:atomic_timeline_simplified}
\end{figure*}

\begin{figure}[ht]
  \centering
  \includegraphics[width=0.8\columnwidth]{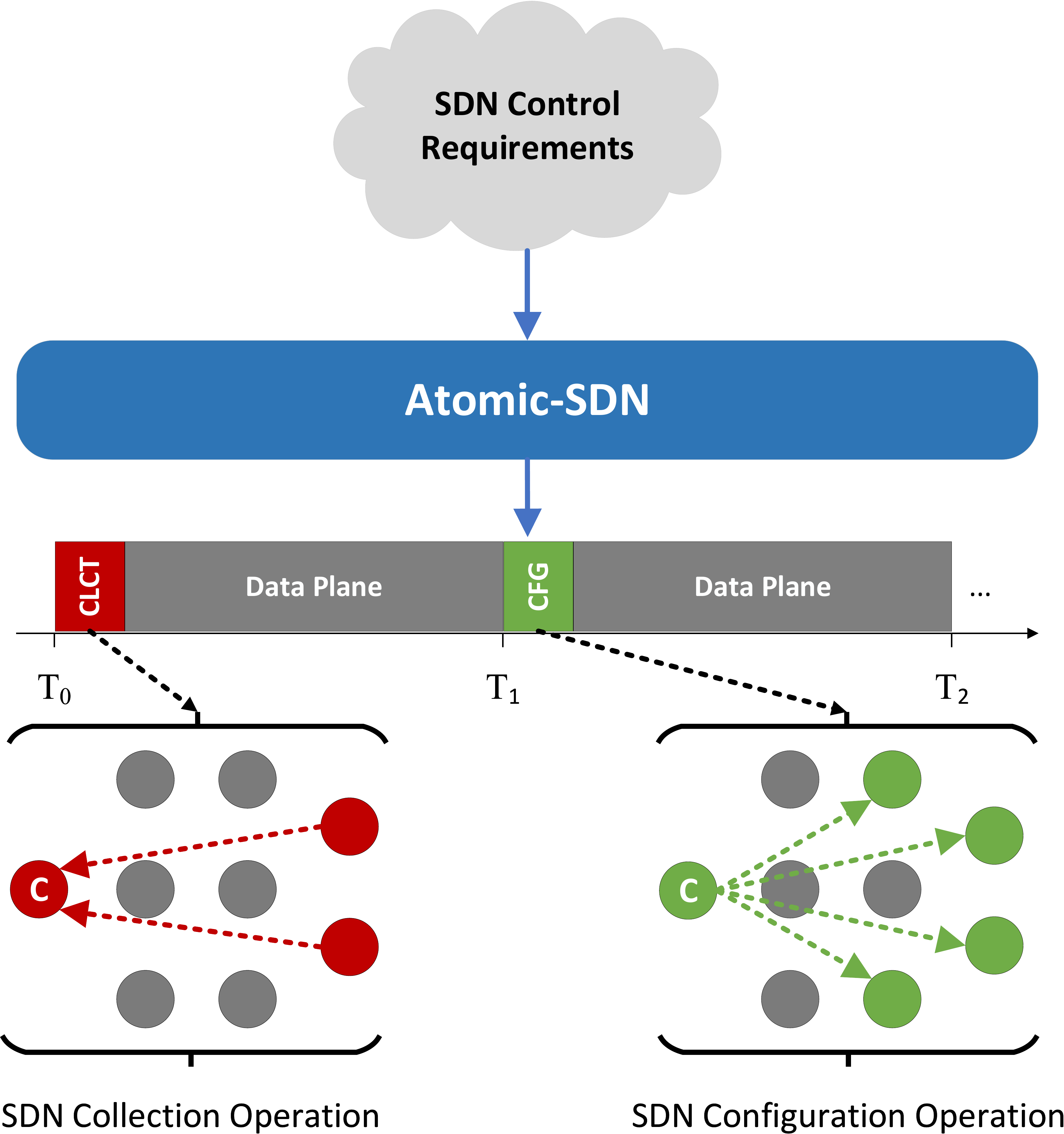}
  \caption{High-level overview of the Atomic-SDN approach in Figure \ref{fig:atomic_timeline_simplified}.} 
  \label{fig:atomic_high_level}
\end{figure}

\noindent\textbf{Motivation:} 
Over the past few years SF has been shown to be extremely capable in delivering fast, reliable communications in low-power wireless networks, and solutions based on this technique have consistently placed within the EWSN Dependability Competition \cite{ewsn-comp-2018-bigbangbus,ewsn-comp-2018-crystalclear,ewsn-comp-2017-robustflooding,ewsn-comp-2017-redfixhop,ewsn-comp-2017-towardslowpower}; achieving high reliability, low latency, and increased energy efficiency over standard approaches. These results support the case for using SF as a platform for SDN control in low-power wireless, the arguments for which are as follows:



\textit{Firstly, the broadcast nature of SF supports \textit{one-to-all} communication within a single flood.} This renders them inherently stateless, without the need for network topology. Currently, multi-hop mesh networks rely on underlying protocols such as Routing for Low-Power and Lossy Networks (RPL), which builds a Directed Acyclic Graph (DAG). This distributed routing protocol is typically used to funnel data from sensor networks towards a single border router, where they are later processed. However, \textit{one-to-many} \textit{downwards} communication is a common issue in RPL networks. An example of this challenge is shown in the \textit{configuration} scenario in Figure \ref{fig:sdn_all_opprtunities}. In this case, the SDN controller wishes to set a Point-to-Point (P2P) link from S$\rightarrow$D across multiple branches of the RPL DAG. The tree-like topology forces the controller to navigate multiple branches to reach all destinations, resulting in packet duplication as it individually transmits to each child. This is particularly relevant in RPL non-storing mode which doesn't support multicast forwarding, although recent efforts attempt to address this \cite{constrained_cast}. This issue isn't specific to SDN in low-power wireless, however the complex requirements of SDN control means it is a highly visible and present issue for SDN implementations based on IEEE 802.15.4 networks.

\textit{Secondly, SF protocols benefit from minimal latency and extremely high reliability.} They are able to aggressively and concurrently propagate across the entire network within short time period. Nodes don't need to worry about interfering with their neighbors and wait for their next transmit opportunity, but are able to immediately relay a packet on a different frequency. This approach, as employed in Atomic-SDN, is shown later in Figure \ref{fig:atomic_channel_hopping}. Additionally, the inherent nature of flooding means that the network benefits from a great deal of spatial diversity. In many scenarios this allows messages to skirt around interference hot-spots without re-transmitting, saving on energy and decreasing latency. This combination of frequency and spatial diversity is particularly relevant when operating over the 2.4GHz band which, in close proximity to external IEEE 802.11 devices, can pose significant interference to low-power wireless networks.

\textit{Finally, the time-synchronized nature of SF allows SDN control to be decoupled from other network processes}. Currently, as neighboring nodes within a low-power wireless network share a single link, any additional control messaging increases contention over scarce resources, causing increased delay and reduced reliability for other control protocols (such as RPL and 6LoWPAN) and application data. Current approaches have tried to mitigate the burden of this overhead through reduction in the number of control messages, the use of source routing headers, and optimization of the protocol \cite{sdn-wise, coral-sdn, usdn}. However, this limits the effectiveness of the SDN architecture: sacrificing responsiveness and fine-grain configurability for performance. SF provides a highly reliable means of communicating to all network nodes, and completely isolating this overhead to free-up network resources.

\noindent\textbf{Approach:}
Atomic-SDN introduces a novel architecture to allow dynamic configuration of SF protocols in response to changing SDN requirements, and slices network resources in time to isolate the SDN overhead from other network processes and temporally decouple the SDN control plane. This allows configuration of other low-power wireless network layers: such as IEEE 802.15.4, Bluetooth Low Energy (BLE), or 6TiSCH. 

Figures \ref{fig:atomic_timeline_simplified} and \ref{fig:atomic_high_level} show a high-level overview of this concept. Atomic-SDN designates these flooding periods as SDN control `\textit{opportunities}', where each opportunity is preceded by an indicator (IND) flood that informs the network of the type of SDN function that will follow (if any), and how that function will be configured: the role of each node, duration of each flood, etc. Using SF in this way allows Atomic-SDN to ensure maximum utilization of network resources within any given control period, with minimal latency and extremely high reliability. Although other flooding solutions match SF in terms of reliability, such as the Asynchronous Flooding (AF) approach used in Bluetooth Mesh, these tend to have far higher latencies in a Radio Duty Cycling (RDC) environment \cite{silabs_ble_mesh_performance}.

Although SF in and of itself is not novel, its application in solving the problem of SDN control in low-power wireless has not yet been explored. We show that, by utilizing SF to service the multiple traffic patterns required in SDN control, reliable and scalable SDN for low-power wireless networks can be achieved within the local mesh network (up to 100 nodes). As control messages are rapidly propagated across the entire network and reliably received in a single flood by all participating nodes, this dramatically reduces the burden of SDN control overhead that has frustrated current approaches. 

\noindent\textbf{Contribution:}
This paper makes the following contributions:

\begin{itemize}
\item We propose SF as a mechanism for facilitating SDN control in low-power wireless networks.
\item We devise a flexible middleware system for the design, instantiation, and scheduling of SF protocols.
\item We apply this solution to the challenge of SDN control in low-power wireless networks, and present Atomic-SDN: a scalable solution that offers considerable improvements in reliability, latency, and energy efficiency over current architectures.
\item We implement Atomic-SDN in Contiki for motes with TI MSP430F1611 Microcontroller, and CC2420 radio.
\item We evaluate Atomic-SDN against other SDN implementations for IEEE 802.15.4, through simulation on emulated hardware.
\item We evaluate Atomic-SDN on a real-world testbed, and show that it can provide considerable reliability and latency improvements over current approaches to SDN control in low-power mesh networks. 
\end{itemize}

\noindent\textbf{Outline}
The remainder of this paper is presented as follows: in \cref{sec:background} we provide a brief overview of related works exploring SDN in low-power IEEE 802.15.4 mesh networks, and provide necessary background on information on Synchronous Flooding (SF) and the concept of Concurrent Transmissions (CTs); we present Atomic-SDN in \cref{sec:design}, where we cover design aspects; \cref{sec:characterization} characterizes Atomic-SDN performance through analysis of theoretical bounds on latency; in \cref{sec:evaluation} we evaluate Atomic-SDN through simulation and compare it to non-flood based SDN architectures for low-power wireless networks; and we conclude in \cref{sec:conclusion}.

\begin{figure*}[t]
\centering
  \begin{subfigure}[t]{0.24\textwidth}\centering
    \includegraphics[width=1\textwidth]{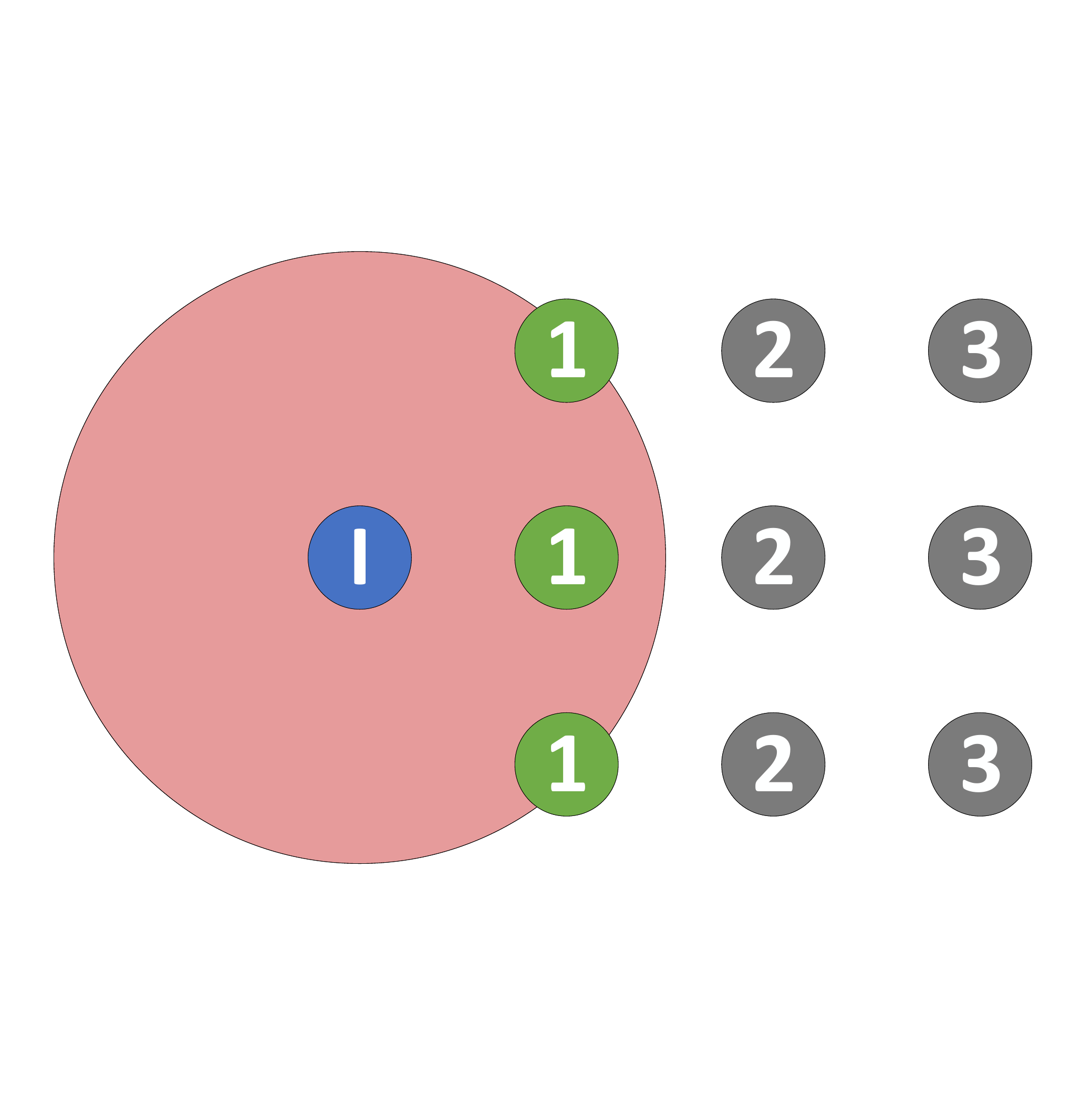}
  \end{subfigure}
  \begin{subfigure}[t]{0.24\textwidth}\centering
    \includegraphics[width=1\textwidth]{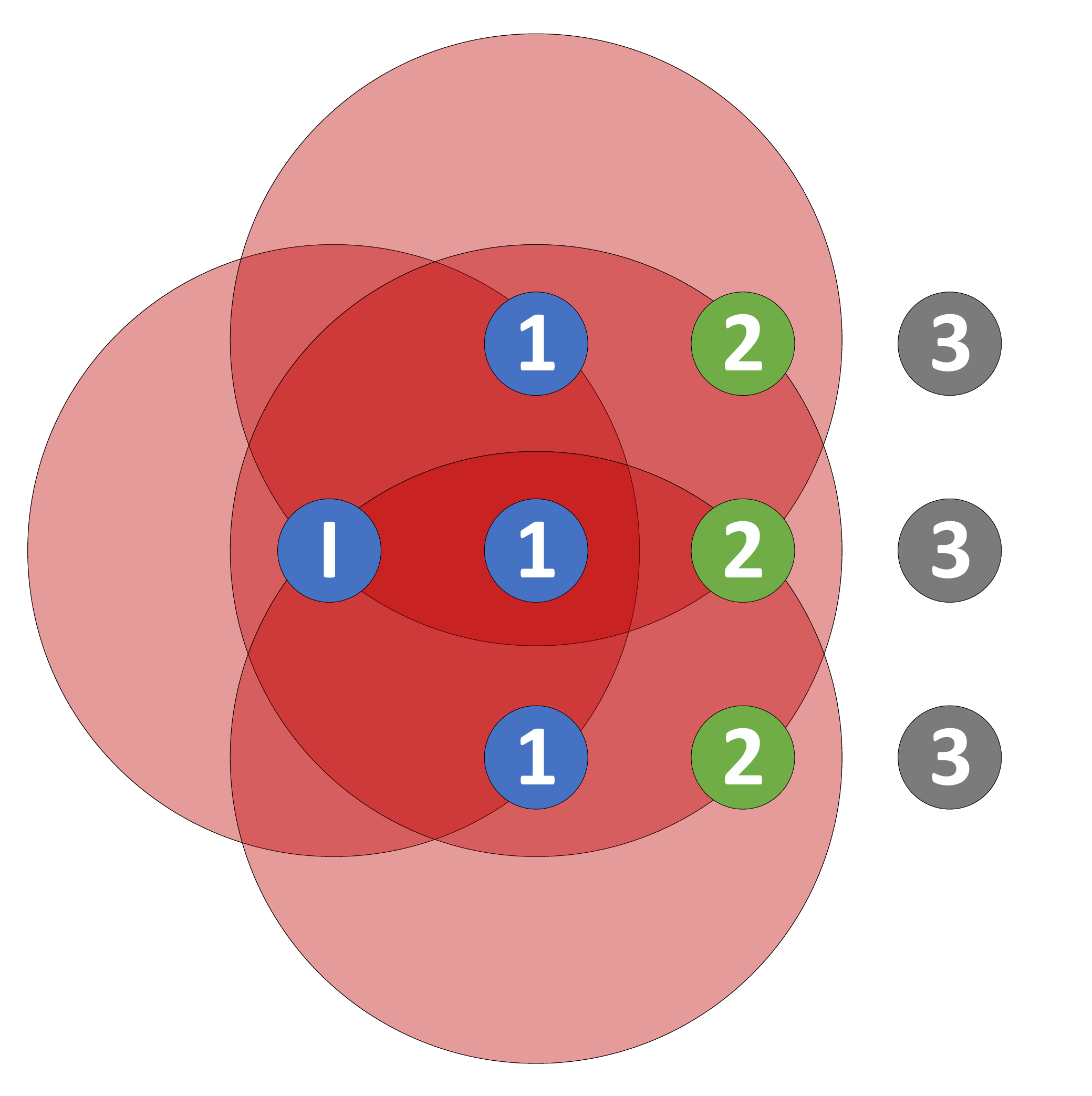}
  \end{subfigure}
  \begin{subfigure}[t]{0.24\textwidth}\centering
    \includegraphics[width=1\textwidth]{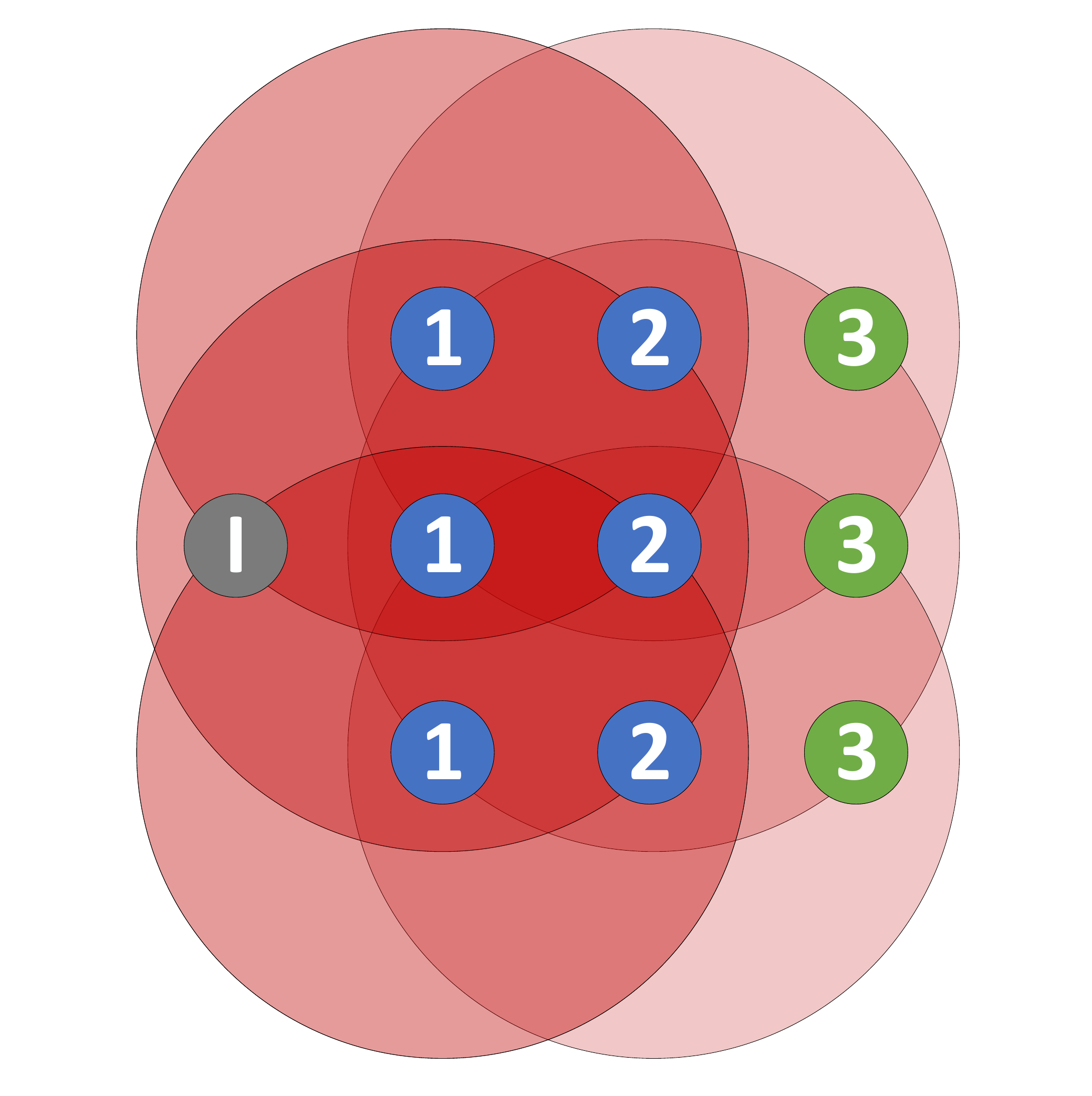}
  \end{subfigure}
  \begin{subfigure}[t]{0.24\textwidth}\centering
    \includegraphics[width=1\textwidth]{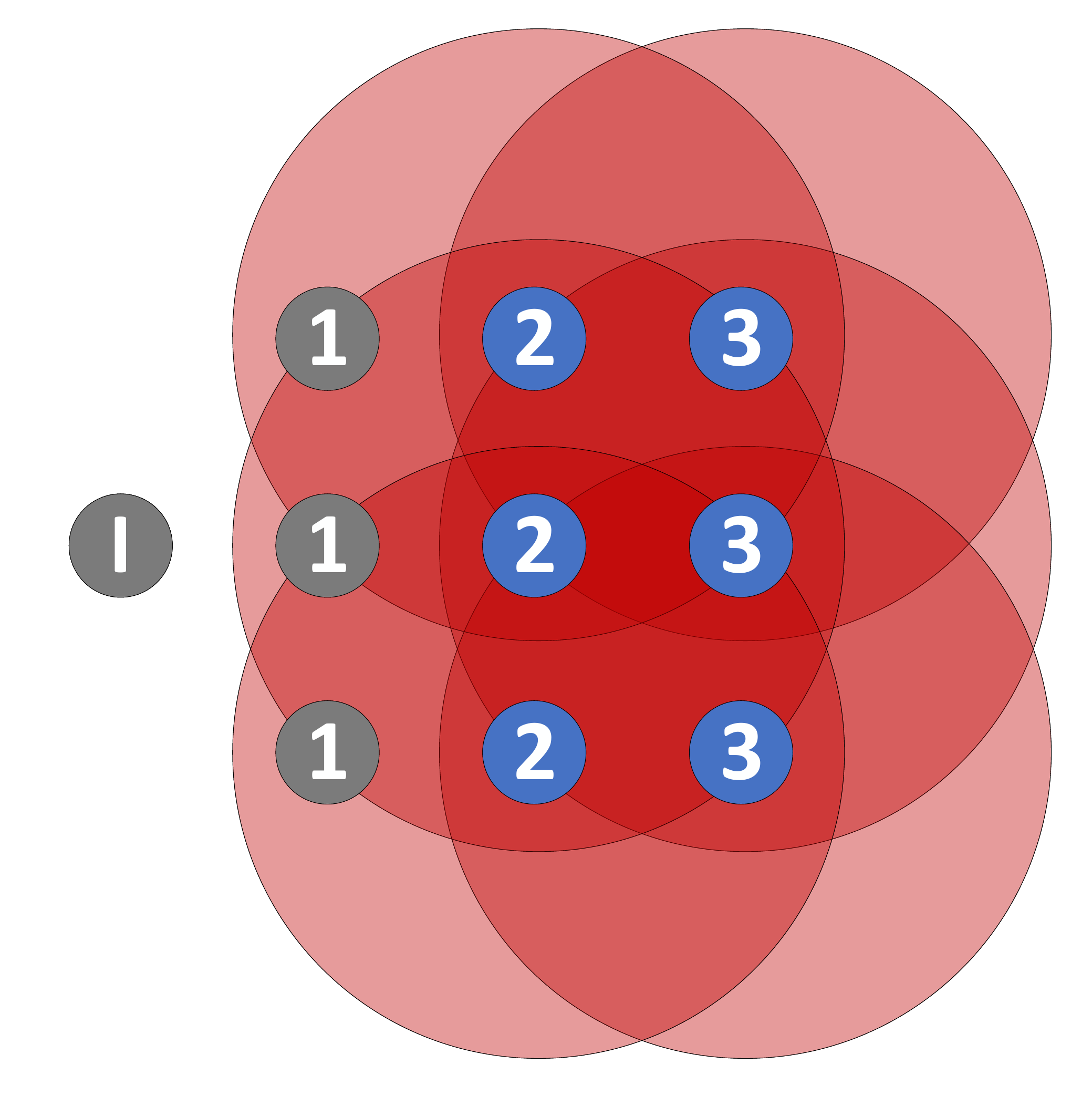}
  \end{subfigure}
\caption{Synchronous flooding using back-to-back transmissions in a 3-hop network (based on the schedule in Figure \ref{fig:flooding_schedule}). Blue indicates a transmission, whilst green indicates a reception}
\label{fig:flooding_topology}
\end{figure*}

\begin{figure}[ht]
  \centering
  \includegraphics[width=1\columnwidth]{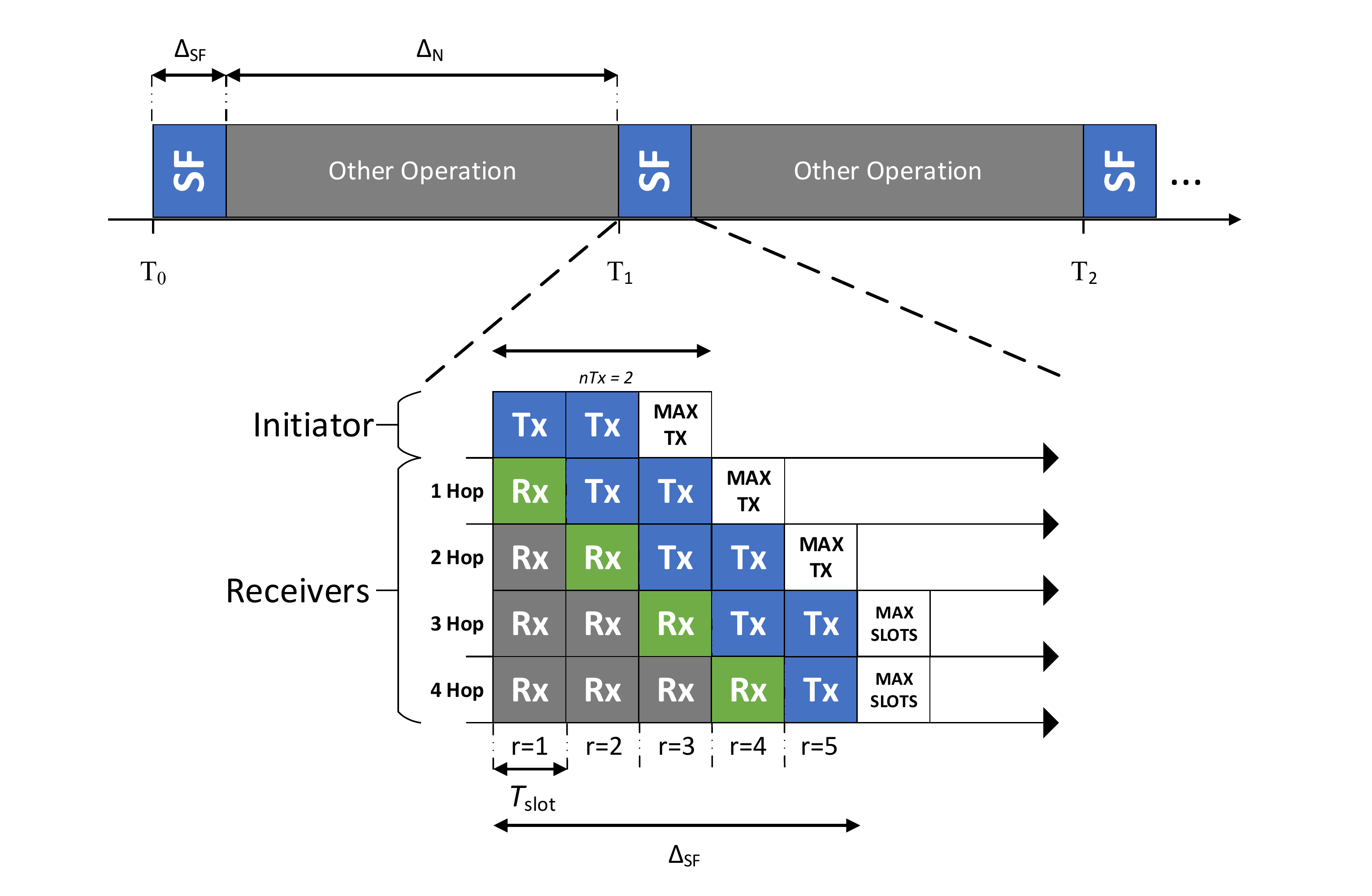}
  \caption{Synchronous flooding protocol used in Atomic-SDN. Back-to-back transmissions flood the network with minimal latency.} 
  \label{fig:flooding_schedule}
\end{figure}

\section{Background Material and Related Works}
\label{sec:background}
This section introduces current approaches examining SDN in low-power wireless networks, as well as necessary background material on Concurrent Transmissions (CT) and Synchronous Flooding (SF).
\subsection{SDN in Low-Power Wireless Networks}
\label{sec:sdn_in_low_power_wireless}
Recent research has considered how to extend SDN control to low-power wireless networks. Whereas traditional SDN concepts have been successfully applied to other networking environments, such as data centers and optical, the constraints of low-power wireless networks (IEEE 802.15.4 in particular) pose considerable challenges to centralized control architectures. We provide a brief outline of current approaches, which have been covered in detail in recent surveys \cite{wsdn_survey_taxonomy, sdwn_opportunities_challenges,sdn_for_iot_survey}, and highlight how they attempt to overcome the challenges of implementing SDN within a constrained environment.

Sensor OpenFlow \cite{sensor-openflow} argues for the use of SDN in sensor networks, proposing a custom low power protocol based on the traditional southbound protocol for SDN, OpenFlow \cite{openflow}. The authors highlight the difficulties of implementing Out-Of-Band (OOB) control plane communication within a sensor network and attempt to mitigate SDN overhead through the introduction of Control Message Quenching (CMQ) \cite{cmq}, whereby retransmissions of SDN control messages from individual nodes are throttled in order to give the controller time to respond to the initial asynchronous request for instruction.

SDWN (Software Defined Wireless Networks) \cite{sdwn} provides an architectural framework and highlights novel uses for SDN in low-power wireless sensor networks. Specifically, the authors introduce the idea of using SDN flowtables to configure in-network data aggregation and Radio Duty-Cycling, allowing the programmatic installation of rules which can help reduce the number of transmissions and improve the energy consumption of individual nodes. In addition, a form of Protocol Oblivious Forwarding (POF) \cite{pof} is proposed to reduce memory footprint, allowing flowtables to match on byte arrays within the packet, rather than needing multiple rules for specific packet types.

SDN-WISE \cite{sdn-wise} builds on architectural concepts introduced in SDWN, as well as introducing stateful flowtables: essentially turning the flowtables into a Finite State Machine (FSM). This allows simple controller logic to be `programmed' into the nodes, where they can perform certain actions under one state, whilst performing a different set of actions when in another. For example, this could be used to allow nodes to run their SDN flowtable actions in a low-energy mode.

CORAL-SDN \cite{coral-sdn} reduces the effect of overhead generated by other control protocols on the SDN stack, and uses a mechanism to reduce RPL control messages in a IPv6 based IEEE 802.15.4 network as nodes initialize and associate with the SDN controller. This frees up resources for the SDN protocol, improving its scalability. However reducing the frequency of RPL control messages may cause issues when trying to maintain end-to-end links between the controller and the edges of the network, particularly in interfered or dynamic networks.

Additionally, SDN concepts are included in recent standardisation efforts from the IETF 6TiSCH Working Group (WG) \cite{6tisch_ietf}, which aims to incorporate elements of SDN within its proposals for centralized scheduling mechanisms. 6TiSCH is engaged in developing scheduling processes for IEEE 802.15.4-2015 TSCH. This amendment allowed the creation of channel hopping schedules but did not define how these schedules should be properly configured or maintained. However, 6TiSCH foregoes traditional SDN elements such as flowtables and focuses the centralized allocation of TSCH slotframe resources (the TSCH {channel/time} slots) within the network.

\subsection{Concurrent Transmissions and Synchronous Flooding}
As previously stated, SF solutions have consistently beaten other approaches in reliability and latency metrics at the IEEE EWSN Dependability Competition. A comprehensive review of SF based protocols can be found in a recent survey and tutorial of CT in IEEE 802.15.4 networks \cite{ci_in_802154_2018}. Through a novel architecture, Atomic-SDN can dynamically configure and support multiple SF protocols from a single framework, allowing SDN control data to be propagated across entire network with minimal latency, while benefiting from high reliability due to the temporal and spatial diversity inherent in broadcast flooding communication protocols. In comparison to contemporary SDN solutions for low-power wireless, this considerably reduces the amount of traffic between nodes and the controller and allows Atomic-SDN to provide a scalable SDN solution in the local mesh (up to 100 nodes).

\noindent\textbf{Temporal Displacement}: The authors of Glossy \cite{glossy_2011} first proposed the use of CT to achieve highly reliable \textit{one-to-many} communication within multi-hop low-power mesh networks. They found that as long as the maximum temporal displacement between concurrently transmitted signals (of the same data) was less than half a microsecond $(\Delta_{max} > 0.5 \mu s)$, then that data can be reliably demodulated without the transmitted signals interfering with one another. 

\noindent\textbf{Managing Clock Drift}: However, the radio-driven nature results in clock drift, meaning synchronization can not be reliability maintained over multiple transmissions. Consequently, the authors proposed interleaving transmission ($T_x$) and reception ($R_x$) slots so that successful receptions could help correct this drift and align the next slot. Recent approaches have demonstrated techniques to estimate this drift \cite{ewsn-comp-2017-robustflooding, raza_competition_2018}, and that slot interleaving is not necessary. This allows nodes to repeatedly $T_x$ after the first reception so that data is forwarded at every slot, meaning the time taken to fully propagate the packet across the network is substantially reduced. This back-to-back $T_x$ approach is utilized as the flooding primitive in Atomic-SDN, demonstrated in Figures \ref{fig:flooding_topology} and \ref{fig:flooding_schedule}.

\noindent\textbf{Flood Operation}: With reference to this back-to-back $T_x$ approach, each flooding period is partitioned into slots. Both the maximum number of slots (MAX SLOTS) and maximum number of transmissions (MAX TX) are statically configured at the start of each flooding round. At the start of each flooding round the initiating (source) node transmits a packet, and repeatedly transmits on every slot until MAX TX. All other nodes have set their radios to receive. The receiving node then relays the packet on next slot, concurrently transmitting with all other forwarding nodes.

MAX SLOTS is used to calculate the maximum flooding time, while MAX TX is the number of times a node concurrently transmits after the first reception. Factors such as external interference, poor connectivity, and the network hop distance need to be taken in to consideration when these variables are set. Increasing them allows for greater reliability, and at a minimum the number of slots needs to equal the hop distance of the network, while minimising them allows for lower end-to-end latency in protocols with multiple flooding periods, as well as reducing energy consumption. In essence, these values are a trade-off between latency and allowing greater temporal and frequency diversity (if paired with slot-by-slot channel hopping \cite{ewsn-comp-2017-robustflooding}). 

\noindent\textbf{Time Synchronisation}:  The length of each timeslot ($T_{slot}$) is determined by the time needed to transmit the packet $(T_{tx})$ (i.e. made up of the preamble, SFD, MPDU, and packet data), a software delay $(T_{sw})$ introduced by the micro-controller, a radio calibration delay $(T_{cal})$, and a processing delay incurred by the receiver radio $(T_{rs})$ incurred by the hardware, s.t. 

\[ T_{slot} = T_{tx} + T_{sw} + T_{cal} + T_{rs} \] 

Non-initiating nodes listen for flood transmissions. When they successfully receive a packet, a relay counter in the header indicates how many hops (and consequently how many slots) have elapsed. Nodes combine this with their knowledge of $T_{slot}$ to calculate the reference time of the initiator and synchronize to that node. After synchronization, they relay the packet on the next timeslot; alongside any other neighbours who also received within that slot. Nodes at the next hop will repeat the process, and so on, until nodes have either reached MAX TX or the flooding period ($\Delta_{SF}$) has elapsed (calculated from $T_{slot}$ and MAX SLOTS).

Once synchronized, the initiating node effectively acts as a timesync for the network, allowing non-initiating nodes to duty-cycle their radio. The benefit of this approach, key to the operation of Atomic-SDN, is that it allows the protocol to be temporally decoupled from normal network operation; allowing it to be run alongside other control and application protocols or, as in the case of Atomic-SDN, be used to regularly configure those protocols.

\noindent\textbf{Multiple Initiators}: Subsequent studies to the original Glossy paper have shown that the receiver is able to reliably demodulate multiple concurrent transmissions of the same data not necessarily because of so-called constructive interference, as the authors first considered, but likely as a result of transmissions being demodulated as non-coherent Minimum-Shift Keying (MSK), as well as Direct Sequence Spread Spectrum (DSSS) minimizing the error rate \cite{murphy_loves_ci, revisiting_so_called_ci, escobar_improving_reliability_latency}. 

When multiple transmitters concurrently transmit different data, then the technique relies upon the Capture Effect \cite{capture_effect_in_fm_receivers_1976}. This is found within IEEE 802.15.4 radios and refers to the phenomenon that the strongest signal out of multiple co-channel signals will be demodulated. It occurs either if one of the signals is around 3dB stronger, although this depends on the particular hardware and modulation schemes used, or if one of the signals is received significantly earlier than the other competing signals. Although the signals may still interfere, there is a high probability that one of the transmissions will be demodulated. 

This property has therefore been used to great effect in \textit{many-to-one} SF data collection protocols in IEEE 802.15.4 \cite{crystal_2016}, allowing multiple initiators to participate in a shared flood. However, there is doubt as to how well these protocols would perform on other physical layers. Indeed, recently the authors of \cite{beshr_multi_hop_bt5} have experimentally demonstrated that when nodes transmit different data and CTs are applied to Bluetooth physical layers (which do not experience capture effects as significantly as IEEE 802.15.4), then reliability drops significantly.

\section{Atomic-SDN Design}
\label{sec:design}

\label{sec:design_general}
Atomic-SDN has been designed to tackle the issues faced by current approaches to SDN in low-power wireless networks. It implements the three core functions necessary for SDN control, as well as providing association with the SDN controller. Moreover, it facilitates these functions \textit{as quickly as possible}, \textit{as reliably as possible}, and maintains \textit{scalability} in the local mesh. These functions and their associated traffic patterns are detailed below.

\begin{itemize}
\item \textbf{Collection (\textit{many-to-one})}: Nodes need to be able to update the controller of their local and neighborhood state, so that the controller can make informed decisions when configuring the network.
\item \textbf{Configuration (\textit{one-to-many}/\textit{one-to-all})}: The controller needs to be able to configure multiple nodes within the network, either to set data flows across the mesh, or to independently provide instruction to a number of nodes.
\item \textbf{Reaction (\textit{many-to-one}/\textit{one-to-many})}: Nodes need to be able to react to unexpected flows or events by soliciting the controller for instruction, and quickly receiving a response.
\item \textbf{Association (\textit{many-to-one}/\textit{one-to-all})}: Nodes need to be able to join the controller and be configured with initial instructions and network settings. 
\end{itemize}

\noindent Atomic-SDN moves away from previous approaches addressing the challenge of SDN architecture in low-power wireless networks. Rather than layering the SDN architecture on top of standard asynchronous or synchronous Layer-2 protocols in the IEEE 802.15.4 networking stack (such as RPL), Atomic-SDN adopts Synchronous Flooding as the mechanism for communication between the SDN controller and nodes within the multi-hop mesh network. 

Indeed, SF is increasingly seen as the `go-to' solution for low-latency control in low-power wireless networks, particularly when applications require highly-robust communication for unpredictable and opportunistic traffic patterns. This view is supported by the consistent and continued success of SF solutions in the IEEE EWSN Dependability Competition \cite{ewsn-comp-2018-bigbangbus,ewsn-comp-2018-crystalclear,ewsn-comp-2017-robustflooding,ewsn-comp-2017-redfixhop,ewsn-comp-2017-towardslowpower}, which benchmarks protocols on reliability, latency, and energy efficiency across multi-hop networks. 
\subsection{General Approach}
Atomic-SDN provides periodic SDN control \textit{opportunities}, where an initial \textit{indicator} (IND) flood instructs all network nodes as to the type of SDN service that will follow (if any), as well as maintaining time synchronization across the mesh (as shown in Figures \ref{fig:atomic_timeline_simplified} and \ref{fig:atomic_timeline}). This allows Atomic-SDN to separate SDN control from the data plane, and slice the network across time so that control messages are no longer in contention with other protocols (such as RPL, 6LoWPAN, or application-layer). Due to the broadcast nature of SF multiple nodes can be quickly and reliably serviced in a single flood, without replicating messages across multiple topology branches. This provides performance improvements orders-of-magnitude over current approaches to SDN in low-power wireless sensor networks.

However, to implement the different core SDN services within a multi-hop mesh network, multiple traffic patterns must be supported (\textit{one-to-all}, \textit{one-to-many}, \textit{many-to-one}, \textit{one-to-one}). Crucially, the plurality of these patterns are not supported by a single SF primitive or protocol and, as such, multiple protocols are needed to fulfil all required communication types. Unfortunately, the complex and low-level nature of SF implementations has meant that, to date, there has been no unified framework allowing multiple SF protocols (such as Glossy \cite{glossy_2011}, Chaos \cite{chaos_2013}, LWB \cite{lwb_2012}, or CRYSTAL \cite{crystal_2016}) to coexist within a single architecture.

Atomic-SDN solves this issue by introducing a novel SF architecture that allows the construction of complex, higher-level communication by applying pre and post logic functions on top of SF primitives. In this manner, different flooding protocols can be configured, instantiated, and scheduled, as the SDN control requirements change; allowing Atomic-SDN to adapt the SF protocol to the SDN service dictated by the controller, and meet application Quality-of-Service (QoS) requirements.

\begin{figure}[t]
  \centering
  \includegraphics[width=1.0\columnwidth]{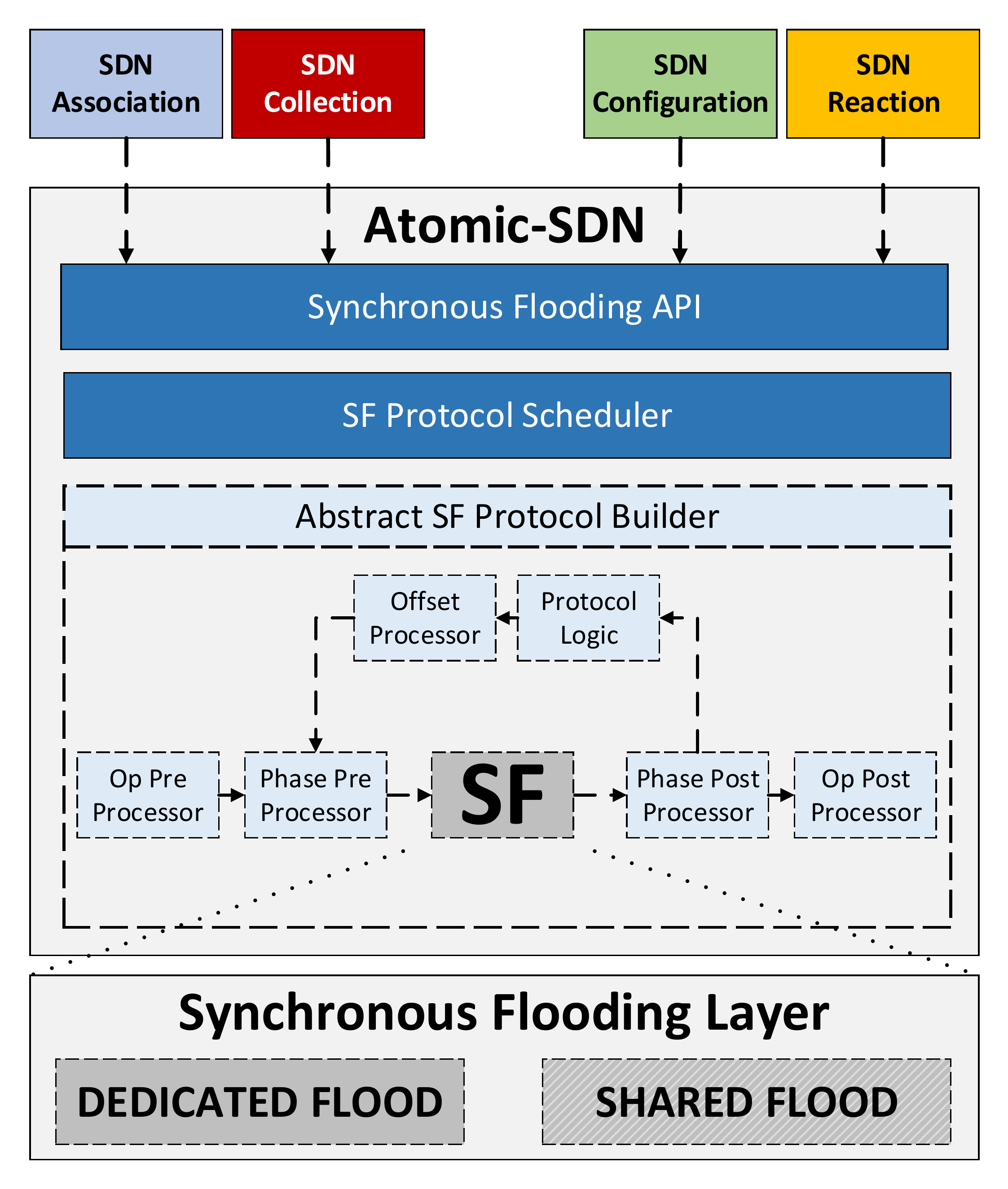}
  \caption{Atomic-SDN architecture.} 
  \label{fig:atomic_architecture}
\end{figure}

\begin{figure}[t]
\centering
  \includegraphics[width=1.0\columnwidth]{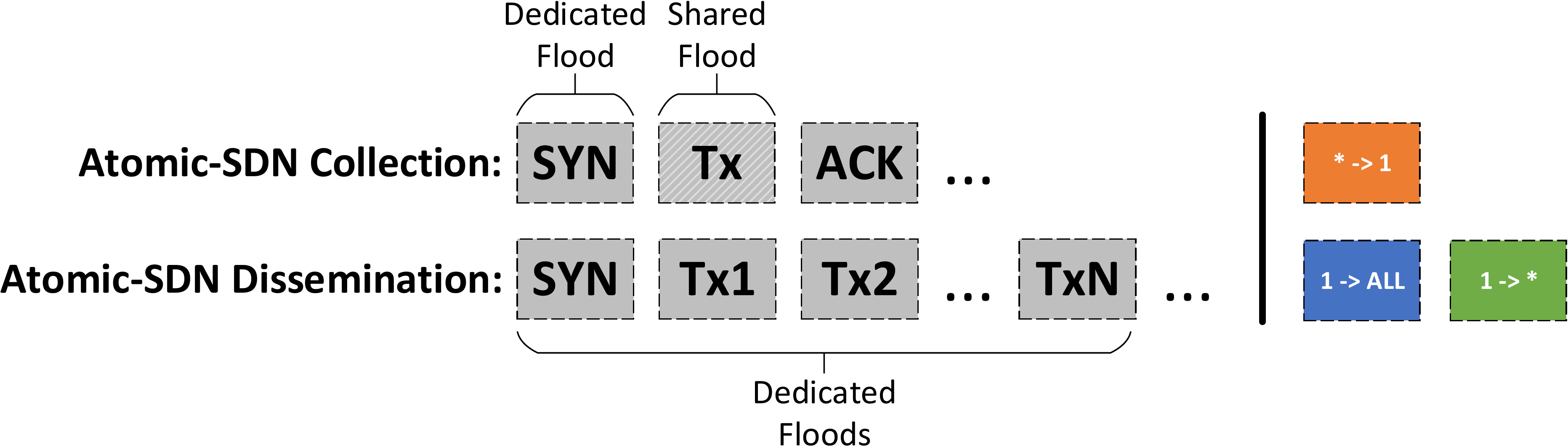}
  \caption{Atomic-SDN data collection and data dissemination protocols.}
  \label{fig:atomic_flood_schedules}
\end{figure}

Figure \ref{fig:atomic_architecture} shows an overview of this architecture. By applying this configurable logic on top of generic flood primitives, SF protocols can be dynamically reconfigured at each SDN control period. The basic approach is as follows (going upwards from the lower layers): 
\begin{itemize}
    \item SF layer manages the lower level time synchronisation and concurrent transmissions.
    \item Floods are packaged into generic dedicated (single initiator) or shared (multiple initiators) flooding primitives.
    \item These primitives are configured with \textit{offsets}, \textit{guards}, and \textit{logic blocks} to create logical \textit{phases}.
    \item Phases are linked and scheduled to create an SF protocol.
    \item SF protocols are mapped to SDN functions, and tailored to their current service requirements, to create an SDN control \textit{opportunity}.
    \item The SDN controller periodically initiates a required SDN operation during a scheduled SF control slot.
\end{itemize}

\subsection{Atomic-SDN Flooding Operations}
\label{sec:design_flooding}
To achieve the core SDN functions; \textit{collection}, \textit{configuration}, \textit{reaction}), as well as network \textit{association}; Atomic-SDN needs to perform three distinct traffic patterns: 

\begin{itemize}
    \item Single source to all destinations (\textit{one-to-all})
    \item Single source to a subset of destinations (\textit{one-to-many})
    \item Multiple sources to a single destination (\textit{many-to-one})
\end{itemize}

\noindent Atomic-SDN provides two SF protocols, (\textit{collection} and \textit{dissemination} shown in Figure \ref{fig:atomic_flood_schedules}), which can be used individually or in conjunction, to fulfil these patterns. Each schedule of dedicated or shared floods repeats until the SDN opportunity is complete.

The first two traffic patterns (\textit{one-to-all} and \textit{one-to-many}) can be achieved through an SF dissemination flooding protocol. In its most simple case, this allows the controller to rapidly and reliably communicate information to the entire network within a single flood, allowing SDN to bypass the packet duplication issues inherent in other SDN architectures for low-power wireless. The flood is then propagated across the network as nodes successfully receive the packet and start to relay the transmission. If a node is designated as a destination (designated in the same manner as before), it will read the packet data after the flood has ended, otherwise it will act as a forwarder. 

The third pattern (\textit{many-to-one}) is more complex, as previously discussed in \cref{sec:background}. In SF collection protocols based on shared flood phases, multiple sources will compete as initiators. In each flood, only one source will successfully be received by the destination. Therefore, competing nodes that were not successful must continue to re-transmit until they are acknowledged in an ACK flood. As nodes are acknowledged they will switch their role from source to forwarder, and help with future transmission phases. This continues until all source nodes have had their transmissions acknowledged, which is indicated by a STOP consisting of one or more empty $T_x$ floods plus a NACK phase.

\begin{figure*}[ht]
\centering
  \includegraphics[width=0.9\textwidth]{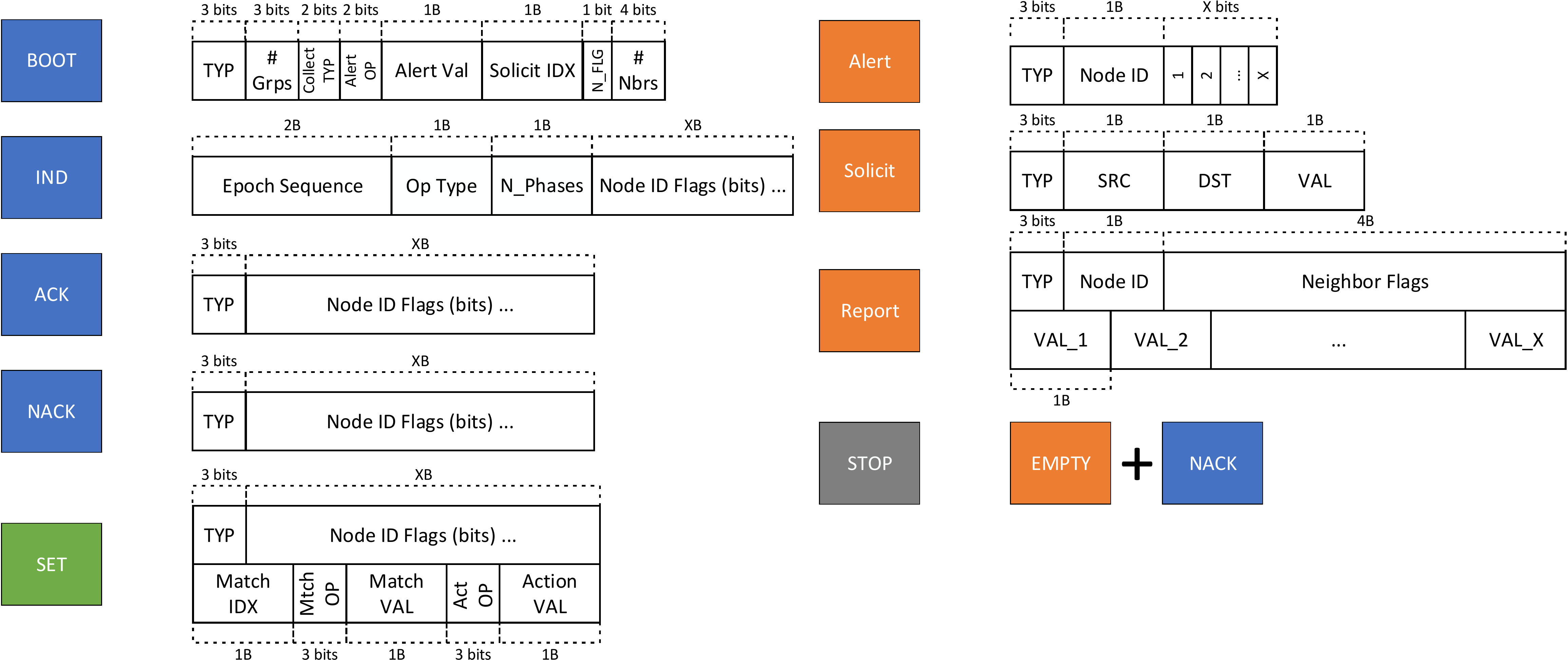}
  \caption{Atomic-SDN \textit{phase} types built from the SF protocols outlined in Figure \ref{fig:atomic_flood_schedules}. These phases are chained together to create higher-level functionality in the form of an SDN `opportunity'. From left to right: \textit{one-to-all} phases (blue), \textit{many-to-one} phases (orange), \textit{one-to-many} phases (green), and a STOP phase (grey).}
  \label{fig:atomic_phases}
\end{figure*}

\subsection{Abstract Protocol Builder}
\label{sec:design_apb}
The success of SF for control solutions in low-power wireless is rooted in the mechanism's ability to provide low-latency and high reliability even under extremely adverse conditions. As such, there have been a number of attempts to take the core flooding principle, and tailor it to diverse application requirements in order to facilitate protocols for \textit{one-to-all} communication \cite{glossy_2011}, data collection \cite{crystal_2016}, \textit{many-to-many} communication \cite{lwb_2012}, network consensus \cite{chaos_2013,a2_2017}, and interference management \cite{crystal_interference_2018}. 

Each of these protocols satisfies a specific set of application requirements. However, to fully implement SDN a number of different traffic patterns need to be supported, and achieving this therefore requires multiple SF protocols. Yet the underlying low-level implementation of proposed SF protocols have, to date, varied significantly; co-existence of multiple protocols within a single stack is therefore particularly challenging despite being based on the same basic underlying mechanism. 

To address this issue, Atomic-SDN implements an Abstract Protocol Builder (APB) middleware layer (as shown in Figure \ref{fig:atomic_architecture}), which uses generic flooding primitives attached with configurable protocol-specific logic to allow flexible construction of complex high-level synchronous flooding protocols. This mechanism is currently used within Atomic-SDN to implement the data collection and data dissemination protocols in Figure \ref{fig:atomic_flood_schedules}; however, the abstract nature of the APB means that it can be easily extended to implement any SF based protocol in order to suit additional traffic patterns or requirements.

\noindent\textbf{Flood Primitives}: In Atomic-SDN, generic `Flood Primitives' are defined as a single flood as shown previously in Figure \ref{fig:flooding_schedule}, configured with a MAX TX number of transmission slots, each with duration $T_{slot}$. If a node is able to successfully complete all MAX TX transmissions it will exit the flood process, otherwise it will exit at $\Delta_{SF}$, the time taken for all transmission slots to elapse. Flood primitives are currently implemented as a \textit{one-to-all} back-to-back transmission flood, however any lower synchronous flooding layer could conceivably be used, such as the Glossy interleaved $R_xT_x$ approach \cite{glossy_2011}, or a consensus primitive such as Chaos \cite{chaos_2013}.

\noindent\textbf{Phases}: Phases are the building blocks of Atomic-SDN, allowing higher-level SDN functionality to be realized by chaining multiple phases into a series of logic decisions. Each phase is a self-contained unit consisting of a flood primitive configured with MAX TX transmissions and duration $\Delta_{SF}$, combined with an associated data structure and the concrete implementation of the following abstract functions, as shown in Figure \ref{fig:atomic_architecture}, which define phase behavior based on the current node role:

\begin{itemize}
\item Pre and post processing logic.
\item Guard to allow for drift and processing in other nodes.
\item Offset from initial phase reference.
\end{itemize}

\noindent By defining these functions, phases can be configured to perform a specific, self-contained role, whilst propagating the associated phase packet types shown in Figure \ref{fig:atomic_phases}. Multiple phases can then be chained together in order to build up higher level processes, known as \textit{opportunities}, allowing full protocols to be implemented through the combination of a number of simple blocks.

\begin{table}[t]
    \renewcommand{\arraystretch}{1.5}
	\caption{Description of phases shown in Figure \ref{fig:atomic_phases}}
    \label{table:atomic_phases}
	\centering
    \begin{tabular}{ C{1.5cm} C{6cm}  }
    \toprule
      	\bfseries Phase Type & \bfseries Description \\ 
      	\midrule
        BOOT & Current SDN settings for network association. \\ 
        IND & Indicate which \textit{opportunity} type will follow (if any). \\ 
        ACK & Acknowledge receptions at the controller. \\ 
        NACK & Acknowledge no receptions at the controller. \\ 
        SET & Configure an entry in the SDN flowtable. \\ 
        ALERT & Alert the controller that an event has been triggered. \\ 
        SOLICIT & Solicit the controller for instruction. \\ 
        REPORT & Report state information to the controller. \\ 
        STOP & End \textit{opportunity} before the allotted time ($\Delta_{XOP}$). \\
    \bottomrule
    \end{tabular}
\end{table}

\begin{figure*}[t]
\centering
  \begin{subfigure}{\textwidth}
    \centering
    \includegraphics[width=.95\textwidth]{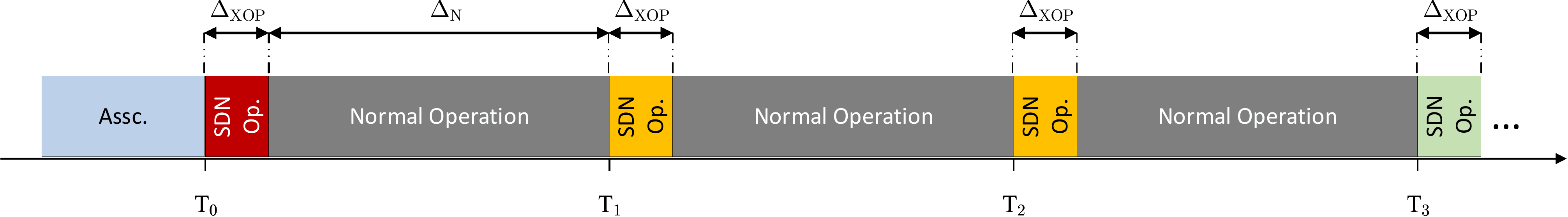}
    \caption{Atomic-SDN epoch schedule.}
    \label{fig:atomic_schedule_1}
  \end{subfigure}
  \begin{subfigure}{\textwidth}
    \centering
    \includegraphics[width=.95\textwidth]{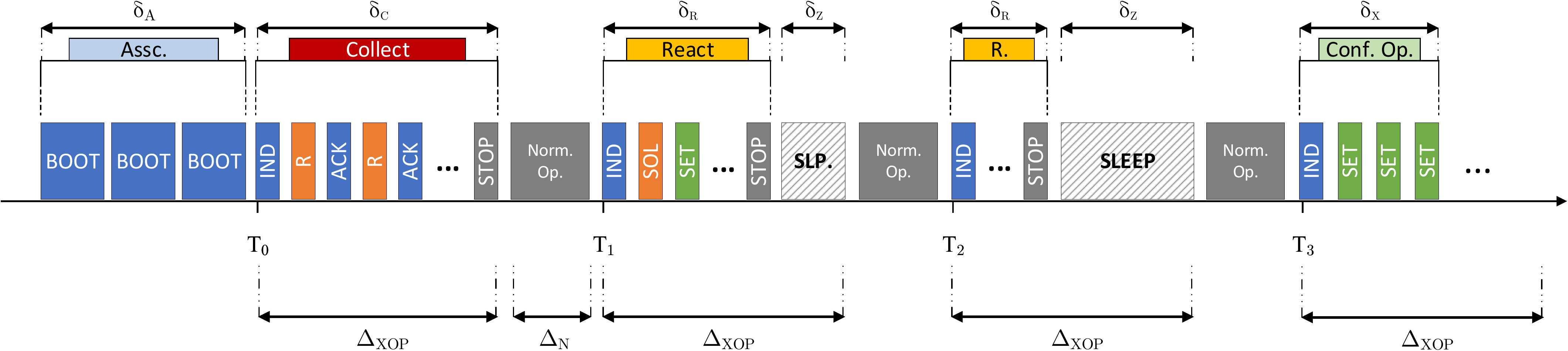}
    \caption{Atomic-SDN opportunity schedule.}
    \label{fig:atomic_schedule_2}
  \end{subfigure}
  \begin{subfigure}{\textwidth}
    \centering
    \includegraphics[width=.95\textwidth]{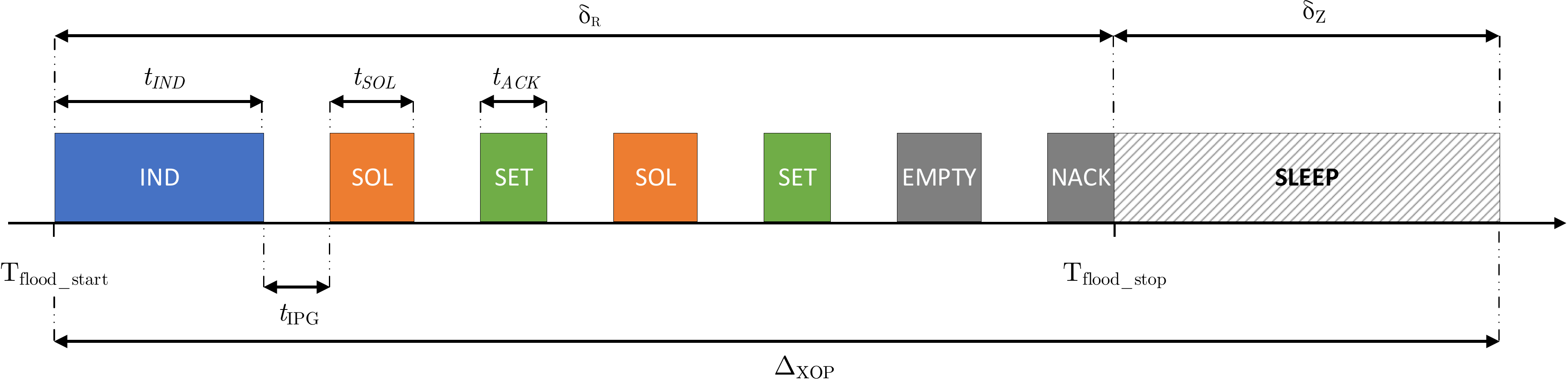}
    \caption{Atomic-SDN phase schedule (REACT Opportunity).}
  	\label{fig:atomic_schedule_3}
  \end{subfigure}
\caption{Example Atomic-SDN schedule at epoch, opportunity, and phase level. Opportunity timings are detailed in \cref{sec:characterization}.}
\label{fig:atomic_timeline}
\end{figure*}

\begin{figure}[ht]
\centering
  \includegraphics[width=.9\columnwidth]{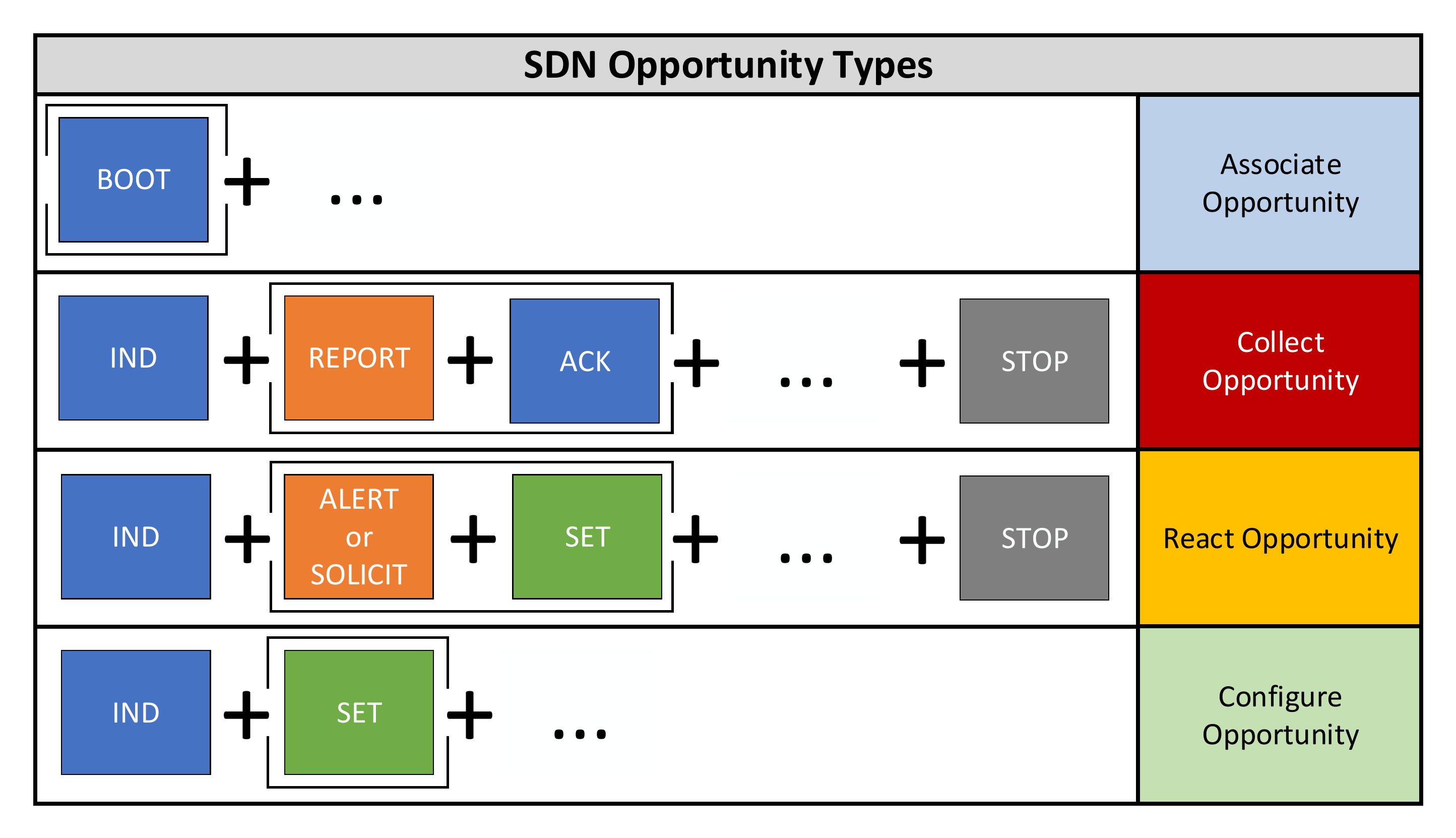}
  \caption{Atomic-SDN control \textit{opportunities} built from the phase types defined in Figure \ref{fig:atomic_phases}. Highlighted phases are repeated until the end of the opportunity.}
  \label{fig:atomic_opportunities}
\end{figure}

\noindent\textbf{Opportunities}: Atomic-SDN defines the concept of SDN opportunities, whereby the controller regularly and synchronously initiates a period of SDN control across the network. These are shown in Figure \ref{fig:atomic_opportunities}, where highlighted phases blocks are repeated until the opportunity is complete; either through a predefined number, or through a STOP phase. The type of opportunity is chosen by the controller prior to the flooding period, where the opportunity logic is constructed through the combination of a number of phase types, along with pre and post processing logic. Prior to execution, every opportunity is announced by the controller through a special \textit{one-to-many} IND phase. This phase instructs the network as to what type of SDN control opportunity to expect (if any), the number of phases in that opportunity, and distributes the current epoch sequence number. Additionally the IND phase includes a variable length array of mapped Node ID flags. Used in conjunction with the current opportunity type, these flags indicate the role of each node within the flood (\textit{source}, \textit{destination}, or \textit{forwarder}).

\noindent\textbf{Epochs}: We define an `Epoch' as the period of time between regularly scheduled SDN control opportunities, with periodicity $T_i$, where a trade-off is considered when setting the epoch length, and consequently the frequency of SDN opportunities. As synchronous flooding periods in Atomic-SDN inherently block other processes, a longer epoch allows a greater amount of time to be devoted to normal network operation; whether that is application processes, other low-power wireless protocols, or to allow nodes to sleep and therefore conserve energy.

\subsection{Scheduling}
Atomic-SDN operates a two-stage scheduling process, as highlighted in Figure \ref{fig:atomic_timeline}. Firstly, self-contained flood `Phases' are chained together within a short period to allow the construction of higher-level SDN functionality. Then, at a macro level, these flooding periods are scheduled periodically to provide regular SDN `opportunities', as well as maintaining tight time synchronization across all nodes.

\noindent\textbf{High-Level `Opportunity' Scheduling}: One of the core principles behind Atomic-SDN is the separation in time of control processes from normal network operation. By slicing control independently from normal operation, the controller is able to define a short period of time in which it is able to communicate with and instruct associated nodes in the local mesh. 

With each control period serving a single SDN control function, this necessitates some decision making and scheduling from the controller: choosing what type of control opportunity to initiate at the start of each epoch, and instructing nodes when to quit the control period and resume normal network operation. This scheduling process is shown in Figure \ref{fig:atomic_schedule_1} which shows a high-level timeline of Atomic-SDN. Once the type of SDN control opportunity is chosen, a mandatory \textit{indicator} (IND) phase is scheduled at the start of the control period. This \textit{one-to-many} phase allows the controller to propagate the opportunity type (if any) to the rest of the network, as well as assigning nodes' roles (\textit{source}, \textit{destination}, or \textit{forwarder}) and distributing any additional information, such as maximum length of the control period.

\noindent\textbf{Low-Level `Phase' Scheduling}: As described in \cref{sec:design_apb}, each flood is packaged into self-contained `phases' which accomplish specific functions within a larger SDN opportunity. After receiving the IND phase propagated at the start of each opportunity, nodes within the mesh participate in a pre-defined schedule mapped to the SDN opportunity defined within the IND, where the schedule consists of a number of distinct phases of one or more types, and each phase (in and of itself) has its own low-level slot schedule. The phase schedule is configured depending on the the current phase type, where guard times, offsets, and protocol logic are determined from the current node role within the context of the larger SDN opportunity.


\subsection{Channel Hopping and Network Association}
Network association is achieved through BOOT and IND phases. BOOT phases distribute the current SDN configuration to joining nodes (match/action information for flowtables, what information should be included in \textit{collect} opportunities, etc.). IND phases are scheduled every epoch and, as well as containing SDN opportunity information, allow nodes to re-associate themselves if they have de-synchronized from the network.  

\begin{figure}[ht]
\centering
  \includegraphics[width=.9\columnwidth]{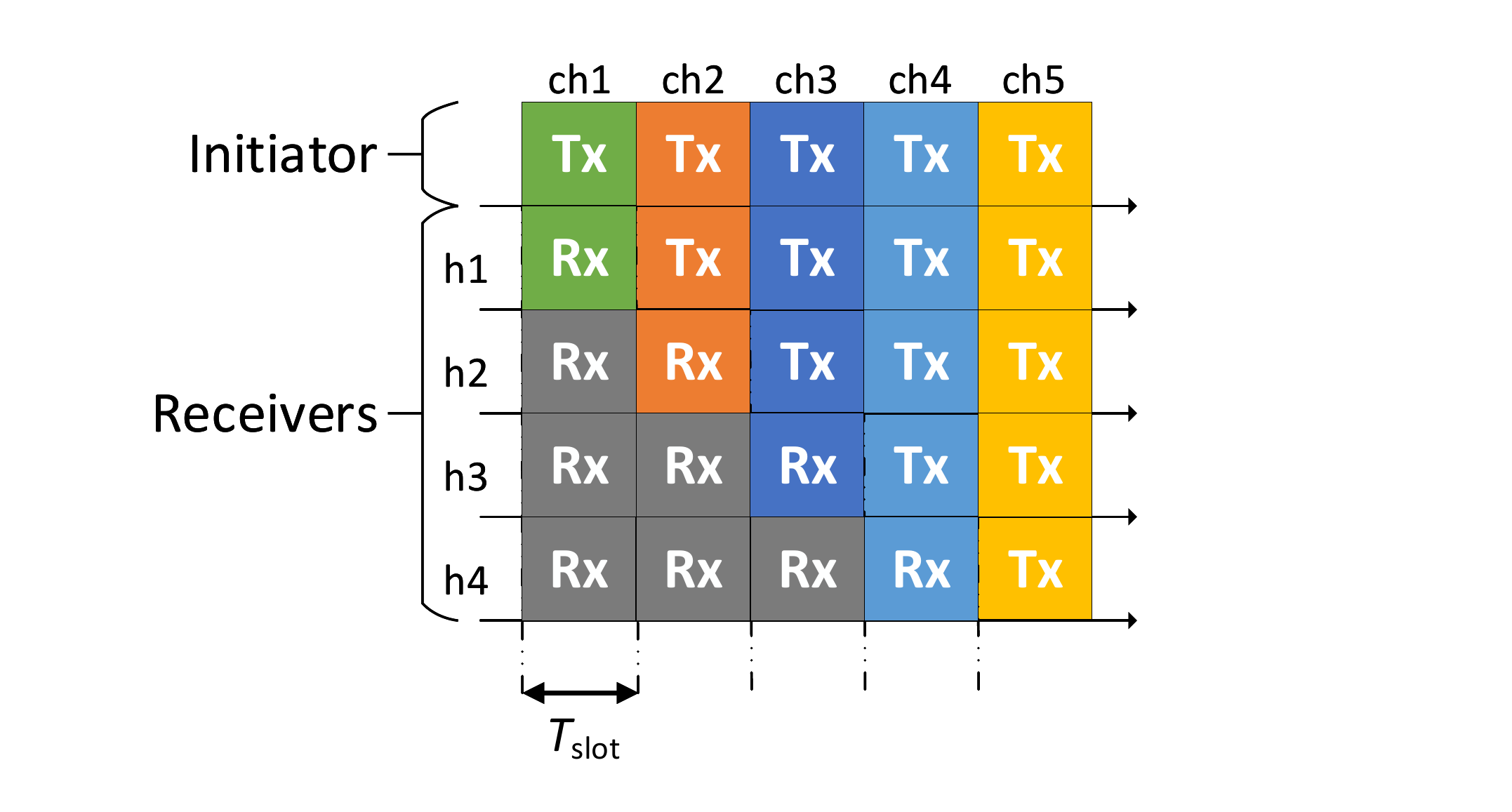}
  \caption{Per-slot channel hopping in an Atomic-SDN flood.}
  \label{fig:atomic_channel_hopping}
\end{figure}

Figure \ref{fig:atomic_channel_hopping} shows how Atomic-SDN employs per-slot channel hopping as proposed in \cite{ewsn-comp-2017-robustflooding}. In every IND or BOOT phase the Atomic-SDN controller distributes the current epoch sequence number, which is used to generate a pseudo-random channel hopping sequence for the network. Once known, nodes increment this number every epoch, meaning that if they miss an IND phase due to their duty-cycling or from interference they will retain knowledge of the hopping sequence. At SF primitive slot, nodes concurrently hop to the next channel in this sequence. Known association channels are seeded into every second channel in the sequence (for example, $\{ch2, ch3\}$), so that when a node is trying to join (or re-join) the network, it merely has to listen on one of these known channels for a long enough period until it hears a transmission and can re-synchronize to the controller.

This mechanism, combined with the spatial diversity of SF, allows Atomic-SDN to survive extremely high levels of interference compared to other SDN architectures for low-power wireless networks, providing reliable network control where other solutions would struggle. It has successfully implemented and tested within competition scenarios \cite{baddeley_competition_2019}, and is evaluated further in \cref{sec:evaluation}.

\section{Characterization of the Latency Bounds}
\label{sec:characterization}
As each SDN opportunity is temporally scheduled, it is possible to characterize Atomic-SDN in terms of the lower latency bounds needed to complete SDN control opportunities (the bounds referenced Figure \ref{fig:atomic_timeline}). This is derived through the following information: \textit{the number of participating active nodes ($n$)}, \textit{the duration of each phase type ($t_{x}$)}, \textit{and the duration of the period between phases ($t_{IPG}$)}.

The time taken in each Atomic-SDN control slice is shown in Equation \ref{eq:2}, where ($\Delta_{XOP}$) is the time spent on the SDN opportunity ($\delta_{SDN-OP}$) plus a variable sleep period ($\delta_{Z}$) up to a maximum control bound dictated by the control periodicity.

\begin{equation} \label{eq:2}
\Delta_{XOP} = \delta_{SDN-OP} + \delta_{Z}
\end{equation}

\noindent ($\Delta_{SDN-OP}$) varies according to the SDN opportunity being run, and depends on the the underlying SF protocol used to facilitate that function.
\vspace{0.2cm}

\noindent \textbf{Configuration}: The configuration opportunity is a \textit{one-to-many} process consisting of an indication phase followed by $n$ number of SET phases and Inter Phase Gaps (IPG), where $n$ depends on the number of individual configuration messages that need to be sent in order to accomplish a specific function within the network. This means that the latency bounds of a configuration opportunity are dependent on the complexity of the function, rather than the scale of the network.

\begin{equation} \label{eq:3}
\delta_{X}(n) = t_{IND} + n*(t_{IPG} + t_{SET})
\end{equation}

\noindent \textbf{Collect}: The collect opportunity is a \textit{many-to-one} process consisting of an indication phase followed by $n$ number of REP/ACK phase pairs and the IPG, alongside the \textit{stop} overhead of 2 empty REP/ACK phase pairs.

\begin{equation} \label{eq:4}
\delta_{C}(n) = t_{IND} + (n+2)*(t_{IPG} + t_{REP} + t_{ACK})
\end{equation}

\noindent \textbf{React}: The react opportunity is a \textit{one-to-one} or \textit{many-to-one} process consisting of an indication phase followed by $n$ number of SOL/SET phase pairs and the IPG, alongside the \textit{stop} overhead of 2 empty SOL/SET phase pairs.

\begin{equation} \label{eq:5}
\delta_{R}(n) = t_{IND} + (n+2)*(t_{IPG} + t_{SOL} + t_{SET})
\end{equation}

\section{Evaluation}
\label{sec:evaluation}
In this section we evaluate Atomic-SDN and compare its performance against $\mu$SDN \cite{usdn}, a publicly available SDN architecture for IPv6 enabled IEEE 802.15.4 networks; and SDN-WISE \cite{sdn-wise}, a SDN implementation for IEEE 802.15.4 based on the RIME communication stack \cite{dunkels2007rime}. Both are implemented in Contiki \cite{contiki}, the same low power Operating System (OS) on which Atomic-SDN is built. Both were chosen evaluation candidates as they can run on top of multiple different IEEE 802.15.4 MAC layers, and therefore provide multiple baselines to compare Atomic-SDN performance. This section demonstrates that, by utilizing SF to create periodic SDN control slices, Atomic-SDN displays considerable performance gains across all metrics in comparison to other SDN architectures for low-power wireless. Furthermore, this mechanism is only possible due to the novel SF framework developed for Atomic-SDN, which allows multiple SF protocols to be configured and instantiated in order to satisfy the plurality of traffic patterns necessary for full SDN control. Figures [\ref{fig:atomic_op_times} - \ref{fig:testbed_pdr_lat}] summarize our results.

\subsection{Simulation and Testbed Setup}
\label{sec:evaluation_setup}
All simulation configuration settings are outlined in Table \ref{table:sim_params}. Simulations were performed using the Cooja simulator and hardware emulator for Contiki OS. Cooja emulates TelosB motes that use the TI MSP430F1611 CPU and CC2420 radio, which is the target platform hardware required by the Atomic-SDN to run the lower layer SF code and is also compatible with SDN-WISE, as well as emulated EXP5438 motes (TI MSP430F5438 CPU and CC2420 radio) used by $\mu$SDN. Additionally, Cooja provides a simulated Multipath Ray-tracer Medium (MRM) radio environment that allows rays to be combined at the receiver (necessary to simulate SF). All simulations were performed across a grid topology, on all nodes, with nodes placed at 300m intervals. 

\begin{table}[ht]
	\caption{Simulation Parameters}
	\renewcommand{\arraystretch}{1.5}
    \label{table:sim_params}
	\centering
    \begin{tabular}{ c c }
    \toprule
      	\bfseries Parameter & \bfseries Setting \\ \midrule
        Duration & 1h \\
        Topology & Grid \\
		Radio Medium & Multipath Ray-tracer Medium (MRM) \\ 
        Link PRR & \textasciitilde90\% \\
        Transmission Range & \textasciitilde300m \\ 
        SNR Reception Threshold & -100dB \\ 
        Capture Effect Preamble & 64$\mu$s \\ 
        Capture Effect Threshold & 3dB \\ 
        SDN \textit{Collect} Period & 60s \\
        SDN Flowtable Lifetime & 300s \\
        Application Data Period & 60$\rightarrow$75s \\
    \bottomrule
    \end{tabular}
\end{table}

As the Atomic-SDN controller needs to keep track of all network nodes, a maximum of 70 nodes were used in simulations due to the memory constraints of the emulated TelosB hardware. The performance evaluation simulations were run over a 1h period, with an SDN opportunity frequency of 1 second. At the start of each opportunity the type of SDN control scenario (either \textit{collection}, \textit{configuration}, or \textit{reaction}) was set in a round-robin process.

Atomic-SDN is evaluated against $\mu$SDN and SDN-WISE, which were configured to adopt two separate MAC scenarios: firstly using ContikiMAC, an energy saving MAC layer, and secondly using \textit{always-on} Carrier Sense Multiple Access (CSMA). In all simulations the controller collects state information from all nodes every 60s, and node flowtable entries have a 300s lifetime. Additionally, each simulation runs a data collection application where nodes attempt to send application data to a sink node at a variable rate of 60$\rightarrow$75s.

Testbed experimentation was performed on a 19 node TelosB testbed. The nodes are located over two floors and exhibit a mixture of dense clusters as well as remote multi-hop branches. As with the simulations, these experiments were also run over 1h periods. However, the focus was solely on examining Atomic-SDN through analysis and evaluation of the SDN \textit{react} operation in an artificially harsh environment, where reception losses of up to 75\% are injected directly into the SF layer.      

\subsection{Simulation: SDN Scalability in the Mesh}
A key challenge for SDN in IEEE 802.15.4 low-power wireless has been to maintain scalability as the mesh grows from a handful, to hundreds of nodes within a local cluster. As the number of nodes increases, the controller needs to appropriate a greater proportion of limited network resources to SDN control. However, in Atomic-SDN the minimum latency bounds discussed in \cref{sec:characterization} establish that certain guarantees can be made concerning the resources needed to complete an SDN function when the number of participating nodes is known.

\begin{figure}[ht]
\centering
  \includegraphics[width=.8\columnwidth]{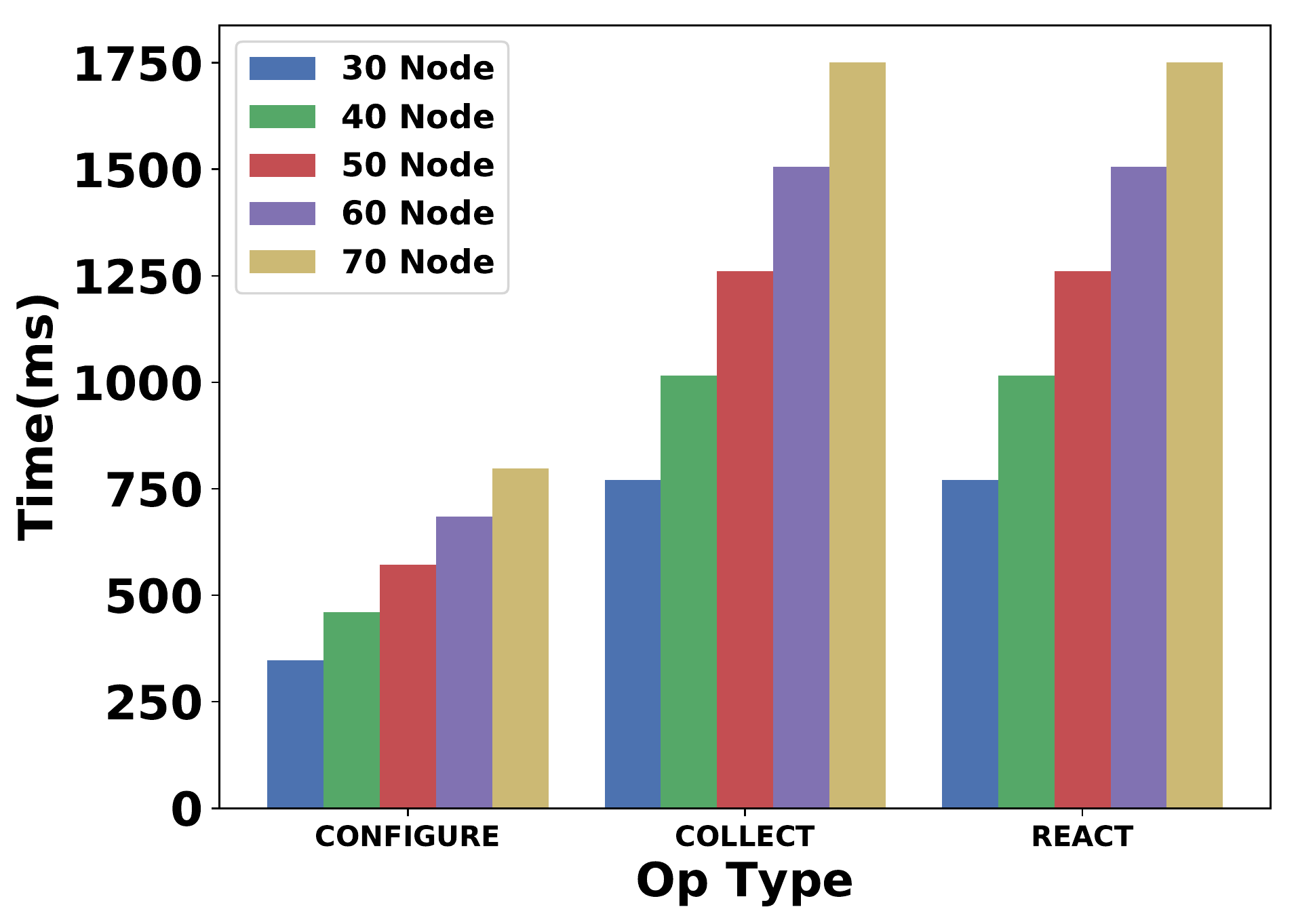}
  \caption{Time taken to complete each Atomic-SDN opportunity as the local mesh scales.}
  \label{fig:atomic_op_times}
\end{figure}

We explore this by evaluating the completion time for the three SDN opportunity types (\textit{collection}, \textit{configuration}, and \textit{reaction}) in networks of increasing size: running simulations for 30, 40, 50, 60, and 70 node mesh networks (limited to 70 nodes due to hardware memory constraints on the Atomic-SDN controller node). Figure \ref{fig:atomic_op_times} shows these results, and demonstrates how the use of SF protocols means that the time to complete all SDN opportunity types (for all network nodes) increases linearly with network size, regardless of hop count. 

In each SDN opportunity type we have assumed a worst-case scenario where the controller needs to interact with each node independently. However the number of nodes participating in each SDN opportunity, necessary to fulfill the requirements of higher-level application functions (virtually located at a centralized controller), would likely be a smaller subset of all nodes and therefore incur lower delay.



The SDN \textit{configure} opportunity is a \textit{one-to-all} process where, after the initial indication phase, each node is configured in turn, in a scheduled fashion. In a 70 node network this allows the configuration of all network nodes within 800ms, assuming each node requires a separate configuration message. However, this time could be substantially reduced (to tens of milliseconds) if a configuration message is relevant to all, or a subset of nodes; allowing the SDN controller to configure multiple nodes in a single flooding phase.

Both the \textit{collect} and \textit{react} opportunities utilize the same underlying SF protocol, and therefore exhibit equivalent delay. In this two phase protocol, the competition between nodes to successfully transmit their data to the controller means that the minimum bound on the completion time is dictated by the number of nodes that need to communicate with the controller. This is a worst-case scenario where it is assumed that all nodes try to perform the SDN operation at exactly the same time, which inevitably causes contention and retransmissions. It does not necessarily follow that this would be the case in a real-world situation. However, these times are still considerably less than the time it takes to complete the same SDN operations in current SDN architectures for low-power wireless networks \cite{sdn-wise, coral-sdn, whisper_programmable_flexible_control, baddeley_evolving_sdn_2018}.

Yet, despite Atomic-SDN demonstrating considerable scalability in comparison to other low-power wireless SDN architectures, there are still questions surrounding the scalability of Concurrent Transmissions when there are 100s or 1000s of nodes and a large number of hops \cite{scalability_of_ci}; particularly in extremely dense networks where the number of concurrent transmitters could be significant. Although the authors of \cite{exploiting_ci_for_scalable_flooding} propose mechanisms for managing these issues and employing CT based protocols across large networks, there has yet to be experimental evaluation of large-scale SF in real-world scenarios.

\begin{figure*}[t]
\centering
  \begin{subfigure}[t]{0.32\textwidth}\centering
    \includegraphics[width=1\textwidth]{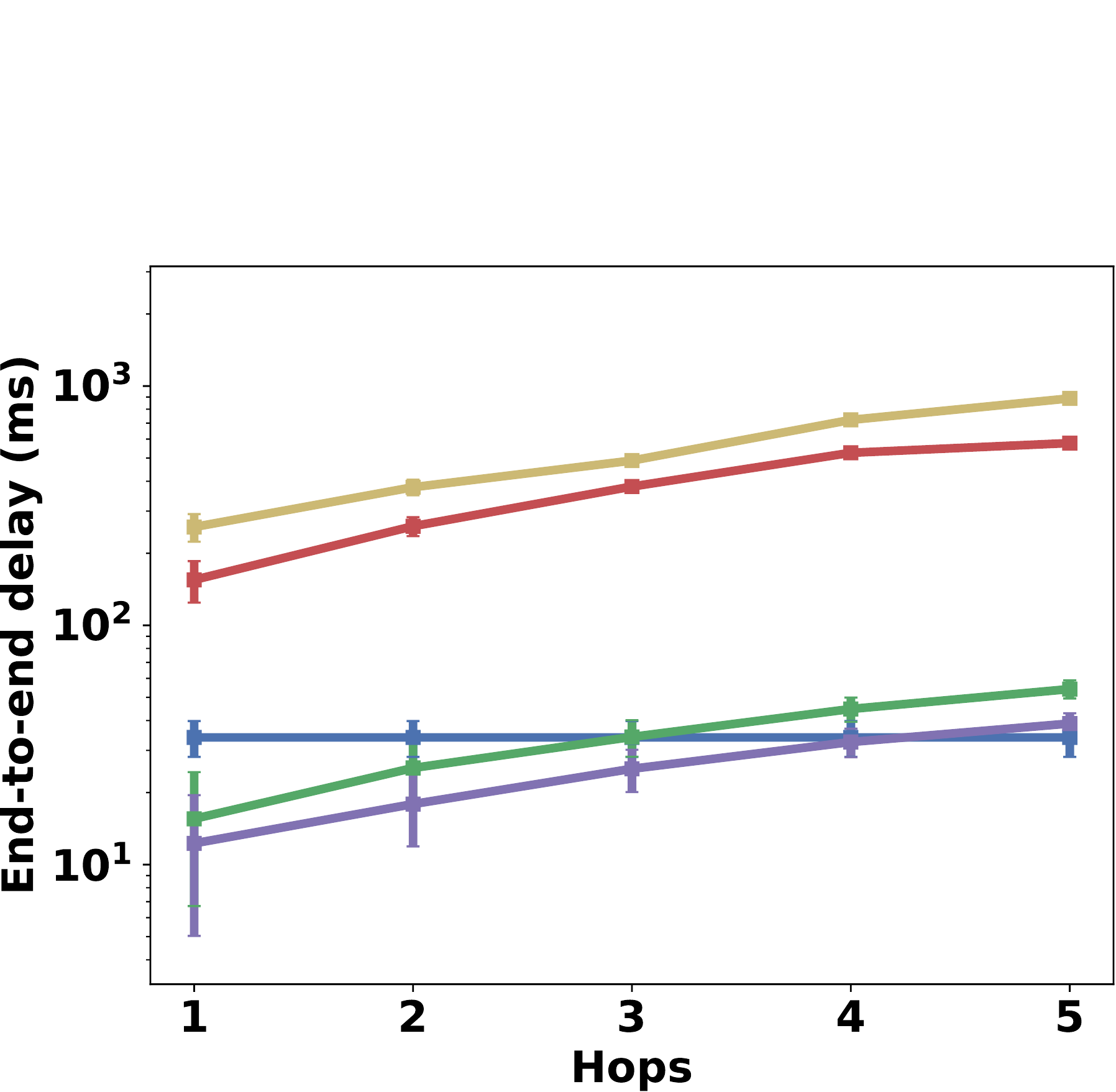}
    \caption{Mean collect delay.}
    \label{fig:collect_lat}
  \end{subfigure}
  \begin{subfigure}[t]{0.32\textwidth}\centering
    \includegraphics[width=1\textwidth]{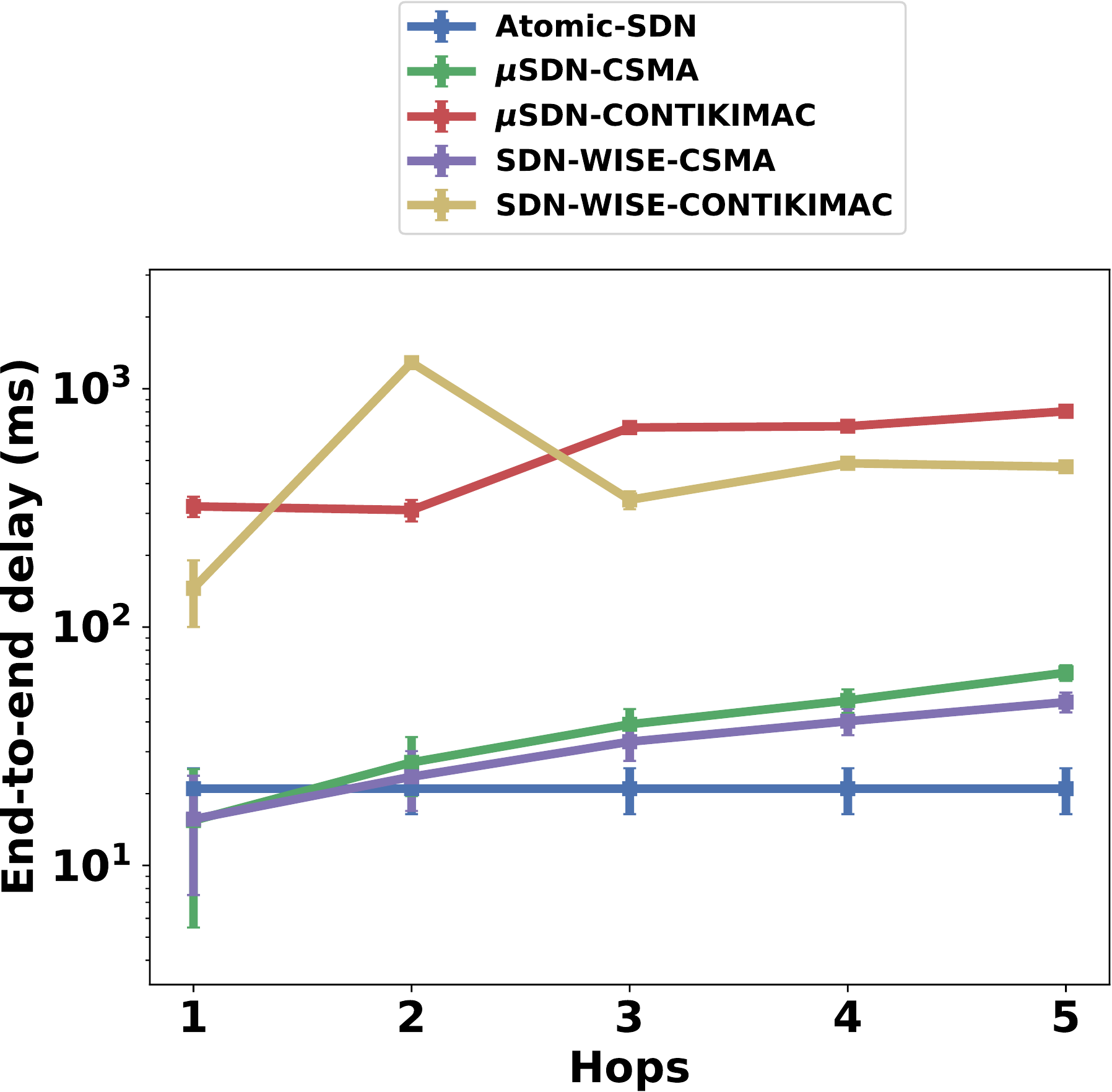}
    \caption{Mean configure delay.}
  	\label{fig:configure_lat}
  \end{subfigure}
  \begin{subfigure}[t]{0.32\textwidth}\centering
    \includegraphics[width=1\textwidth]{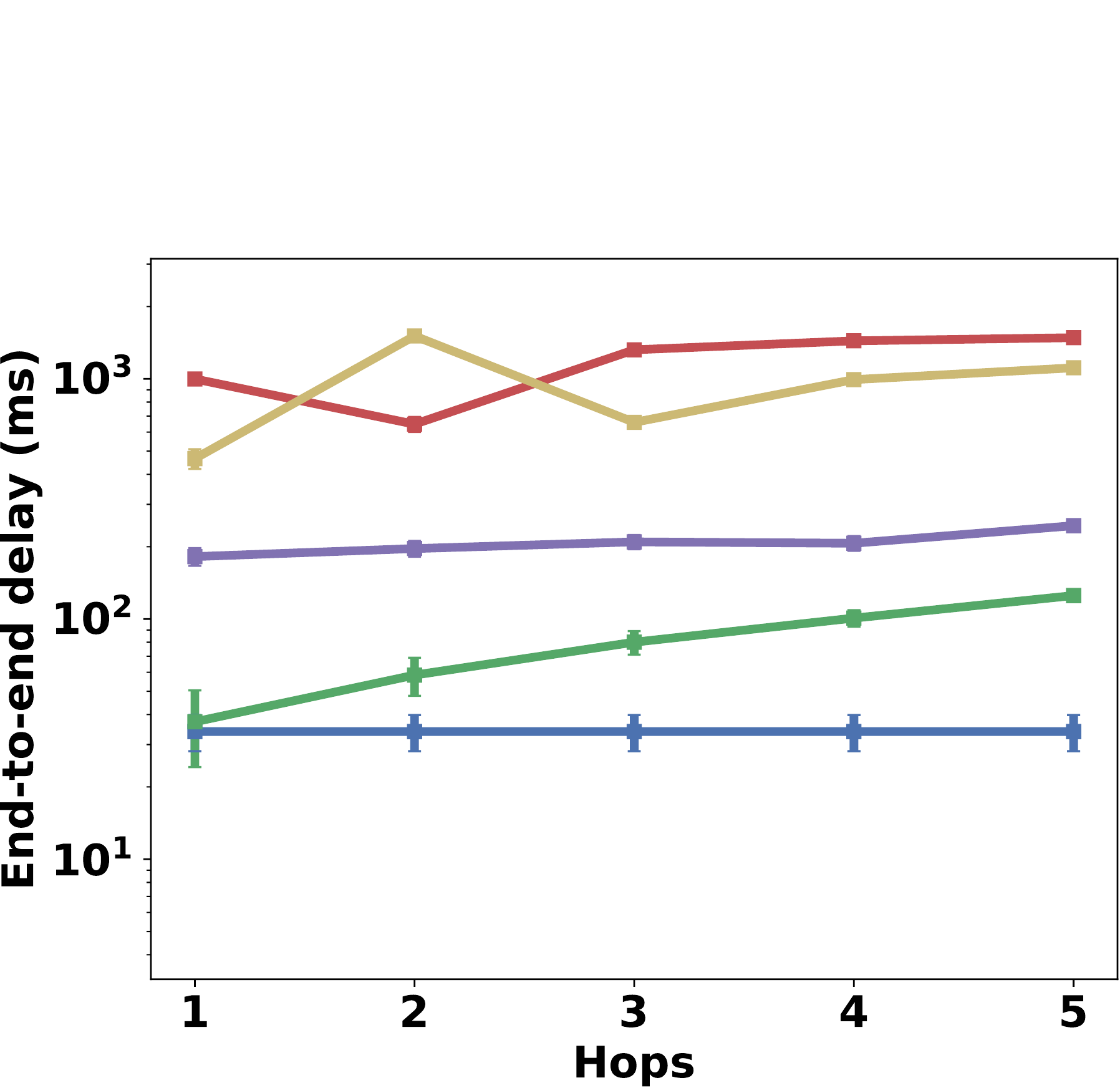}
    \caption{Mean react delay.}
    \label{fig:react_lat}
  \end{subfigure}
\caption{Mean collection, configuration, and reaction delays versus hop distance in 30 node network when an individual node participates. Atomic-SDN exhibits similar latencies to CSMA-based $\mu$SDN and SDN-WISE, however it additionally retains consistent delay across all hop counts.}
\label{fig:best_case}
\end{figure*}

\begin{figure*}[t]
\centering
  \begin{subfigure}[t]{0.32\textwidth}
    \includegraphics[width=1\textwidth]{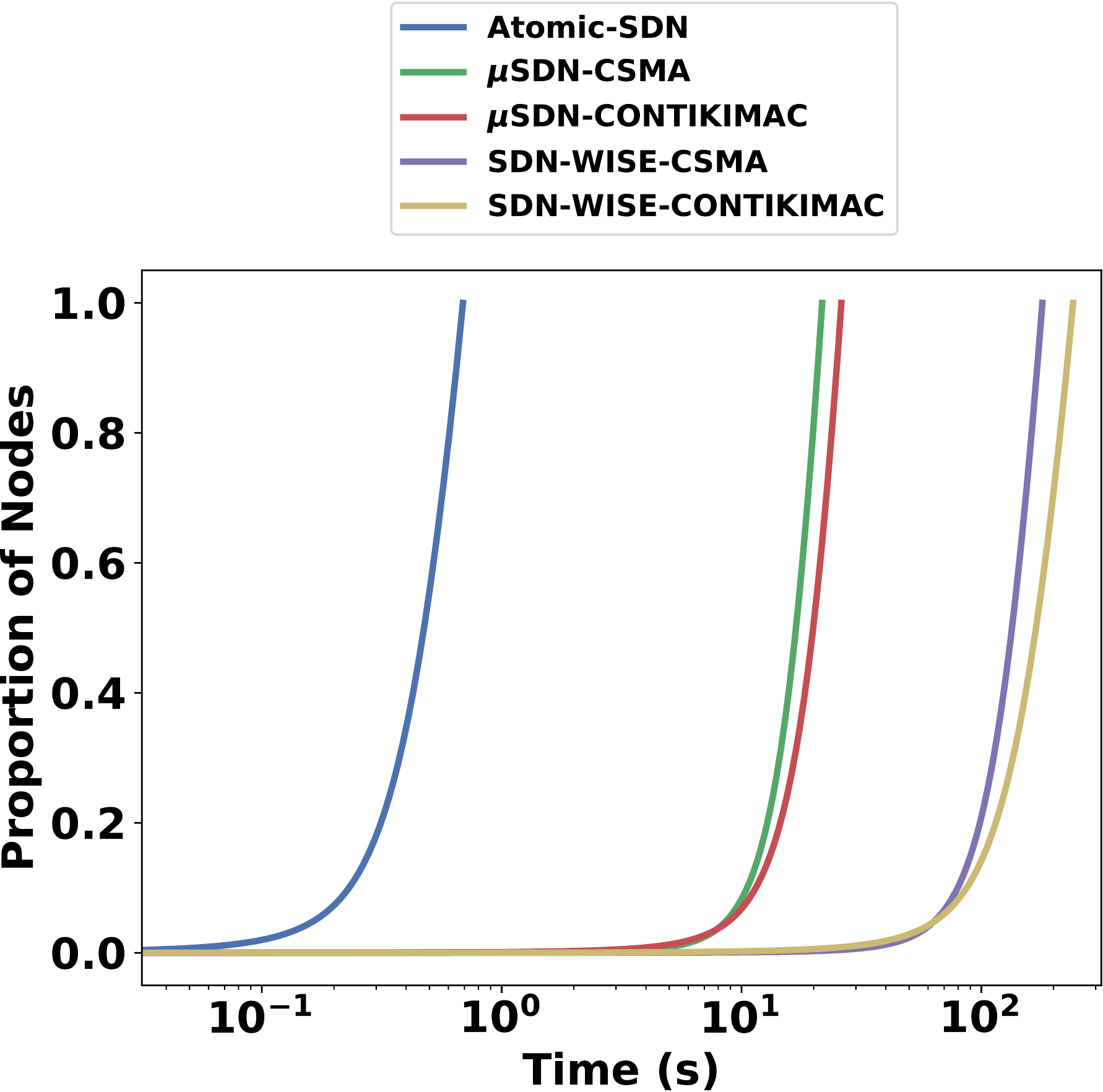}
    \caption{Time taken for an SDN \textit{react} operation.}
    \label{fig:react_latency}
  \end{subfigure}
  \begin{subfigure}[t]{0.32\textwidth}
    \includegraphics[width=1\textwidth]{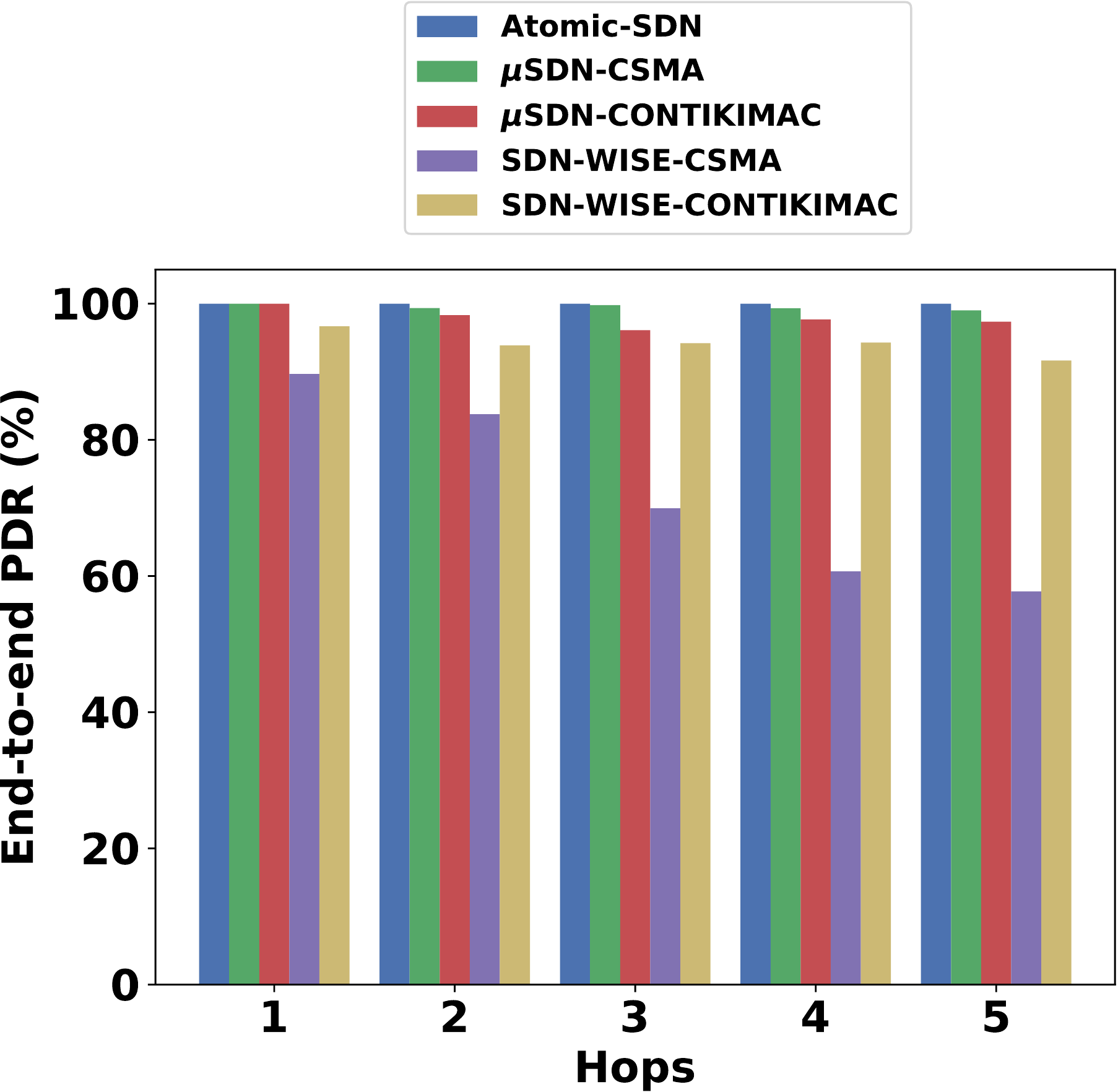}
    \caption{Mean PDR for all SDN control traffic.}
  	\label{fig:pdr}
  \end{subfigure}
  \begin{subfigure}[t]{0.31\textwidth}
    \includegraphics[width=1\textwidth]{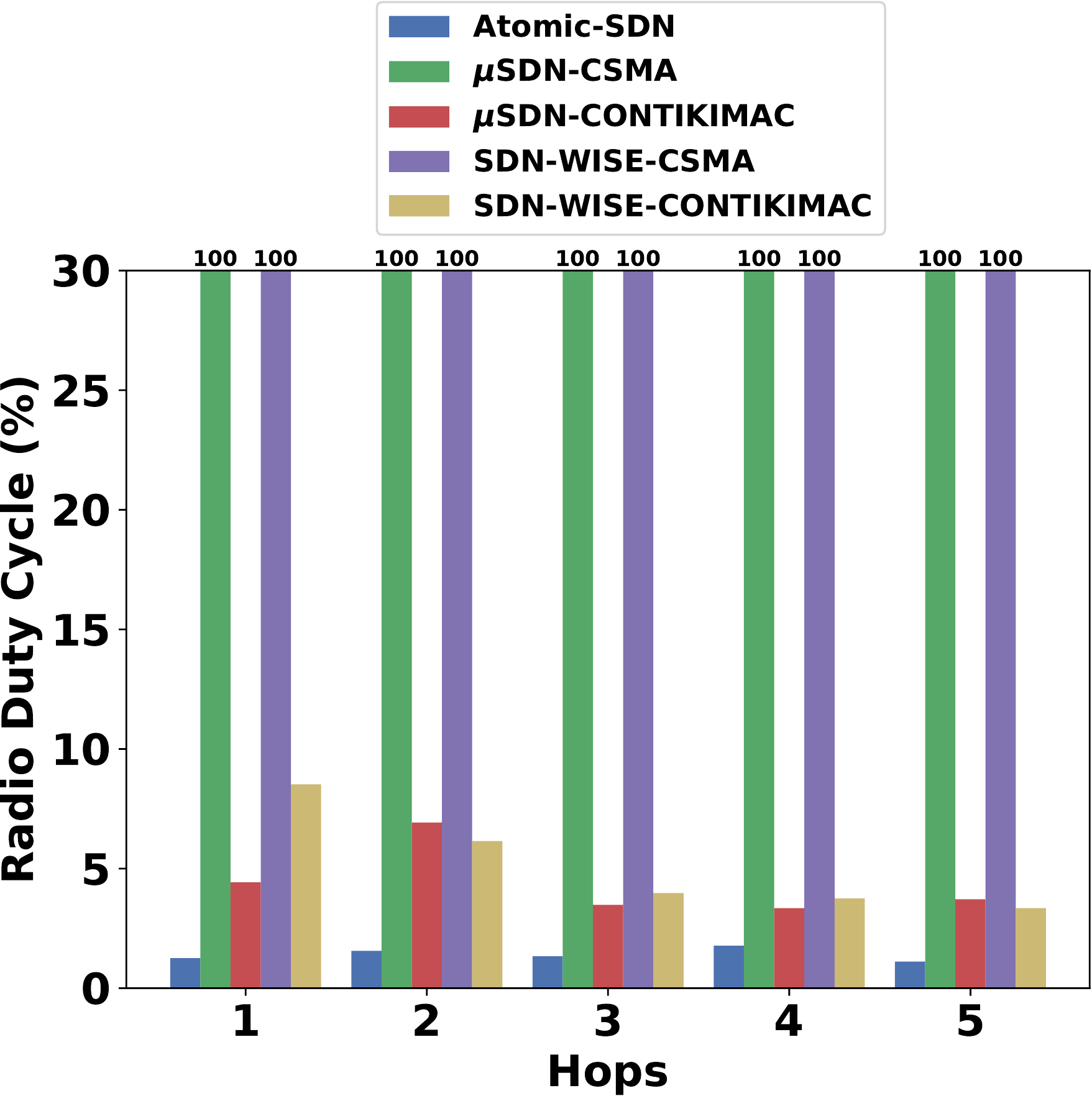}
    \caption{Mean RDC per hop distance.}
    \label{fig:rdc}
  \end{subfigure}
\caption{(a) Time taken to complete an SDN \textit{react} operation concurrently for all nodes in a 30 node network, as well as (b) end-to-end Packet Delivery Ratio (PDR), and (c) Radio Duty Cycling (RDC) versus hop distance from the controller. As CSMA based $\mu$SDN and SDN-WISE are always-on, they exhibit 100\% RDC, denoted at the top of the plot.}
\label{fig:worst_case}
\end{figure*}

\subsection{Simulation: Performance Comparison}
We compare Atomic-SDN against current approaches to SDN in IEEE 802.15.4 using $\mu$SDN and SDN-WISE in two configurations: always-on CSMA and the duty-cycled ContikiMAC. Simulations were performed in Cooja using emulated target hardware, and Atomic-SDN performance gains are evaluated in terms of \textit{latency}, \textit{reliability}, and \textit{energy efficiency}. The simulated network was limited to 30 nodes in order to accommodate SDN-WISE which, although specified as able to support longer routing headers, does not have this feature available in the current SDN-WISE Contiki implementation.

\begin{table}[ht]
	\renewcommand{\arraystretch}{1.5}
	\caption{Mean latency, Packet Delivery Ratio (PDR), and Radio Duty Cycle (RDC) for a single node to perform each SDN opportunity type in a 30 node network.}
    \label{table:performance_summary}
	\centering
    \begin{tabular}{ l c c c }
    \toprule
      	\bfseries Architecture & \bfseries Latency (ms) & \bfseries PDR (\%) & \bfseries RDC (\%) \\
      	\midrule 
        Atomic-SDN & 34.0 & 100.00 & 1.34 \\ 
        $\mu$SDN-CSMA & 33.94 & 99.58 & 100 \\ 
        $\mu$SDN-ContikiMAC & 340.42 & 96.45 & 3.76 \\
        SDN-WISE-CSMA & 25.75 & 68.49 & 100 \\
        SDN-WISE-ContikiMAC & 544.23 & 93.84 & 5.15 \\
    \bottomrule
    \end{tabular}
\end{table}

\noindent\textbf{`Best-Case' - Individual Node Participates:} We first consider a `best-case' scenario where just a single node participates in the SDN control operation (i.e. there is no competing SDN control traffic from other nodes). Table \ref{table:performance_summary} averages these results and Figure \ref{fig:best_case} shows mean delay versus hop distance from the SDN controller for an individual node performing one of three SDN operations: \textit{collection}, \textit{configuration}, and \textit{reaction}. In each case, Atomic-SDN  maintains consistent latency over all distances, due to the minimum bounds on latency inherent in SF protocols. However, as expanded upon in \cref{sec:characterization}, this bound is affected by the underlying SF protocol which supports it (which can require $N$ number of phases). In this scenario, where there is only one individual node communicating with the controller, it is possible for CSMA based architectures to achieve better latency results than Atomic-SDN at lower hop distances depending on the configuration of the lower layer SF primitive (guard times, number of slots, etc.). However, in CSMA, the radio is always on. Compared to the duty-cycled ContikiMAC configurations of $\mu$SDN and SDN-WISE, Atomic-SDN maintains minimal latency bound by the length of the flood, regardless of hop distance.

\noindent\textbf{`Worst-Case' - All Nodes Participate:} We next evaluate performance when considering a `worst-case' scenario where all nodes in the network need to participate in the SDN control operation. These results are presented in Figure \ref{fig:worst_case}.

The SDN \textit{react} operation is used to benchmark the time taken for all network nodes to concurrently solicit and receive instruction a single from the controller. Figure \ref{fig:react_latency} shows the effectiveness of SF protocols in comparison current SDN implementations for low-power wireless, where nodes need to perform two or three-way handshakes across multiple Layer 2/3 links. Not only is Atomic-SDN able to perform this operation on all network nodes within milliseconds, but this is orders-of-magnitude faster than the non-SF approaches, which can take seconds or even minutes.

Figure \ref{fig:pdr} shows Atomic-SDN achieves 100\% reliability compared to both the CSMA and ContikiMAC configurations of $\mu$SDN and SDN-WISE. As $\mu$SDN implements end-to-end acknowledgements for SDN control traffic, it presents with a higher Packet Delivery Ratio (PDR) in comparison to SDN-WISE (particularly the CSMA configuration) which has no transportation layer guarantees. Additionally, ContikiMAC causes high channel utilization through packet retransmissions. Although this improves the overall PDR in the case of SDN-WISE, as packets have a higher chance of surviving each hop, it also causes considerable contention on links that experience high traffic loads. This therefore results in a drop in PDR for $\mu$SDN nodes at greater hop distances as they contend with nodes nearer the controller.

Finally, Figure \ref{fig:rdc} shows the Radio Duty Cycling (RDC) at each hop. ContikiMAC configurations show reduced energy efficiency at lower hop counts as nodes need to serve messages from their children; whilst energy for Atomic-SDN increases at higher hop counts as nodes closer to the controller receive in the first few transmission slots on an ACK, and so spend less time participating in the flood. CSMA configurations, which don't perform any duty-cycling, display high PDR across all hops. Barring any contention, nodes should always be able to receive transmissions as the radio is always on. In comparison, the use of SF in Atomic-SDN means it can benefit from a highly reliable MAC, whilst retaining the low-energy operation of duty-cycled approaches. By using the APB to construct multiple SF protocols tailored for each SDN task Atomic-SDN demonstrates near perfect reliability when collecting state information from all nodes. Although these results are based on simulation and 100\% reliability cannot be guaranteed, multiple studies and extensive experimental evaluations (as covered in a recent survey \cite{ci_in_802154_2018}) have demonstrated up to 99.99\% reliability is achievable using SF protocols, even under heavy interference.

\subsection{Testbed: SDN Control Under Interference}
\label{sec:evaluation_testbed}
\begin{figure}[ht]
\centering
  \includegraphics[width=0.8\columnwidth]{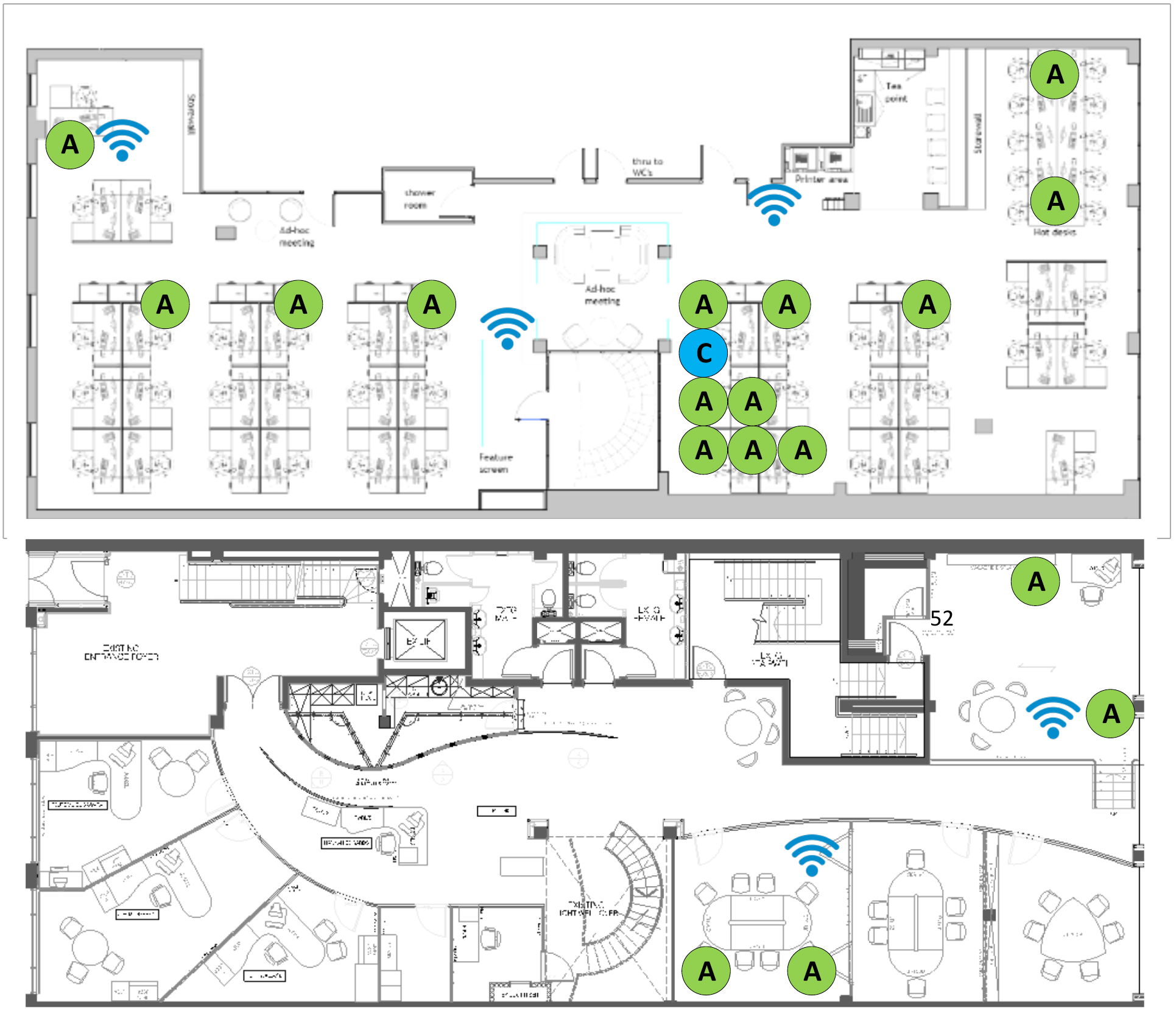}
  \caption{19 Node TelosB Testbed. Initiating nodes are marked in green, and the SDN controller in blue.}
  \label{fig:testbed_layout}
\end{figure}

A 19 node testbed was used to evaluate the potential of Atomic-SDN to provide reliable SDN control in high interference scenarios, reaching all mesh nodes irrespective of link quality. The layout of this testbed is shown in Figure \ref{fig:testbed_layout}: nodes are located over two floors and it presents a number of interesting features such as a dense cluster, isolated multi-hop paths, and non Line-of-Sight (LOS) lossy links.

\begin{figure}[ht]
\centering
  \includegraphics[width=0.8\columnwidth]{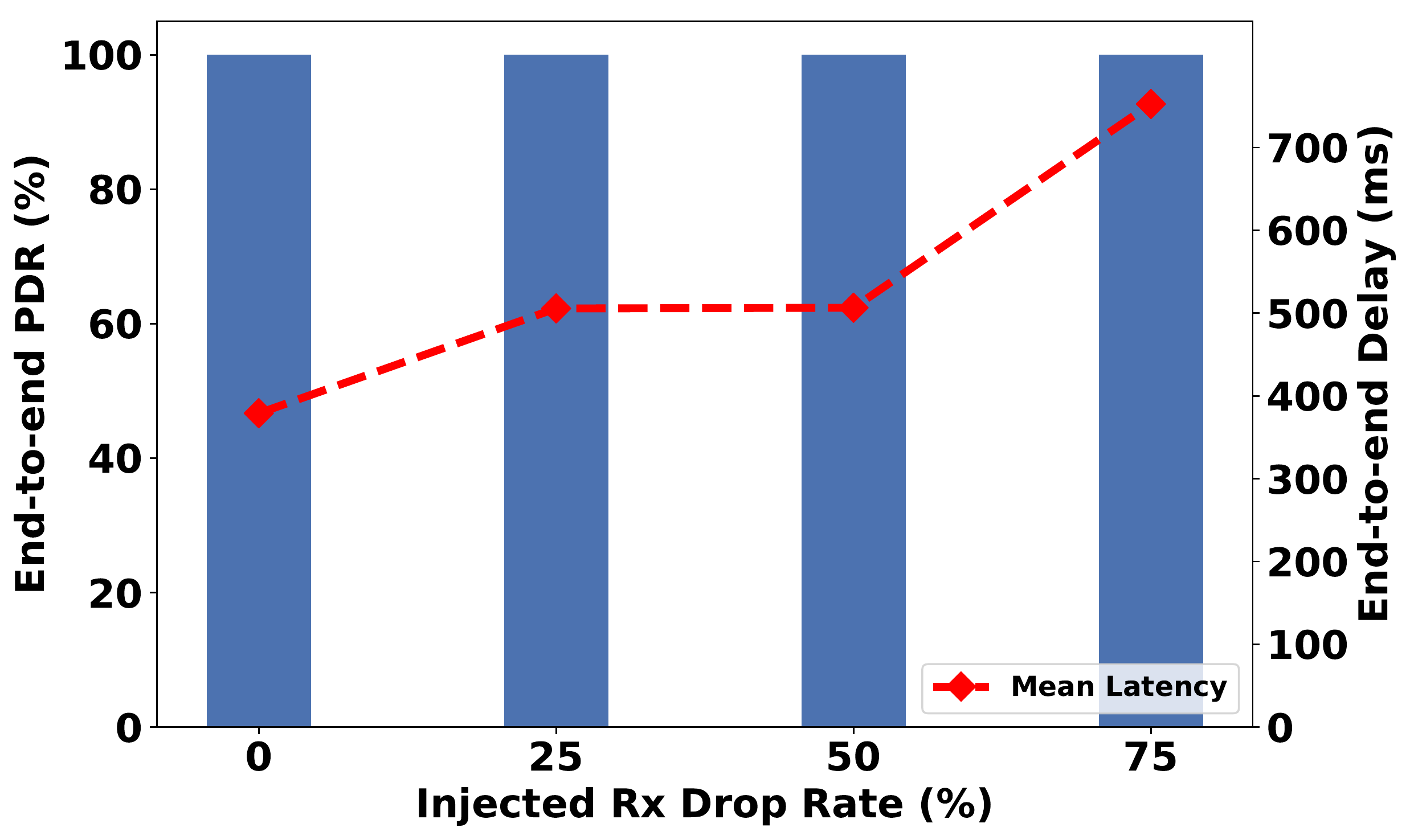}
  \caption{PDR overlayed with mean round-trip latency for SDN \textit{react} operations, injected with probabilistic reception misses, in a 19 node testbed. Atomic-SDN maintains high reliability even during 75\% injection of receive misses.}
  \label{fig:testbed_pdr_lat}
\end{figure}

The SDN controller was configured to repeatedly initiate an SDN \textit{reaction} operation from all nodes with minimal periodicity (i.e. back-to-back) over 1h periods. Forces reception ($R_x$) drops were injected into the SF layer at each node in order to simulate the effects of interference, stepping from 0\%, 25\%, 50\% to 75\% probability. Latency and PDR results for each drop rate are presented in Figure \ref{fig:testbed_pdr_lat}, and show how Atomic-SDN is able to maintain near 100\% probability even with 75\% reception misses. However, it shows that as the probability of missing a reception increases, this also increases the average time taken for each node to complete the SDN operation. As receptions are missed, the underlying SF protocol ensures nodes will re-transmit their controller solicitation data until they hear an acknowledgement. 

Not only do these testbed results demonstrate the resilience of the SF approach to SDN control pioneered in Atomic-SDN (and also shown through success at the IEEE EWSN Dependability Competition), but overcoming this issue of resilience and reliability is particularly relevant in both Advanced Metering Infrastructure (AMI) and industrial control scenarios: where nodes may suffer from lossy links due to harsh environments, or multipath effects due to surrounding industrial machinery and assets. By maintaining a reliable and low-latency link to an SDN controller, based on a stateless and topology agnostic control mechanism, Atomic-SDN can provide programmable SDN control without assuming network stability.

\section{Conclusion}
\label{sec:conclusion}

Attempts to apply SDN concepts within low-power wireless have consistently been met with the same problem: that implementing centralised control architectures for constrained, low-power mesh networks generates complex and considerable overhead that the underlying physical and MAC layers struggle to handle. A number of approaches have managed to reduce this overhead, but a truly responsive and dynamic SDN architecture needs to be able maintain a timely view of the network state, handle network changes within milliseconds rather than seconds or minutes, and be resilient to interference. This is particularly relevant if SDN research in low-power wireless is to move beyond sensor networks and start addressing other aspects of IoT such as Industrial Wireless Control for Cyber Physical Systems, where strict latency and reliability requirements cannot be met by current SDN control solutions. 

Synchronous Flooding, a radically different approach compared to current standards, can provide this. This paper has introduced Atomic-SDN, a unique solution for SDN in low-power wireless networks that utilizes SF to provide highly-reliable SDN control with minimal latency and extremely high reliability, without affecting application-layer traffic or other control processes. By facilitating the propagation of control messages across the network in a flood, within dedicated control timeslots, Atomic-SDN allows the SDN layer to operate without knowledge of topology, as well as benefiting from the spatial and temporal diversity inherent within flooding protocols. Furthermore, Atomic-SDN has demonstrate resilience to extremely high levels of interference, not only through the testbed experimentation in this paper, but also in extensive benchmarking against WiFi interference at the 2019 IEEE EWSN Dependability Competition. 

\bibliographystyle{IEEEtran}
\bibliography{access}

\EOD

\end{document}